\documentclass[journal]{IEEEtran}

\usepackage{graphicx}
\usepackage[table,xcdraw]{xcolor}
\usepackage{lscape}
\usepackage{longtable}
\usepackage{mathptmx}
\usepackage{latexsym}
\usepackage[hyphens]{url}
\usepackage{booktabs} 
\usepackage[colorlinks = true,
            linkcolor = blue,
            urlcolor  = blue,
            citecolor = blue,
            anchorcolor = blue]{hyperref}
\usepackage{xtab, makecell}
\usepackage{amsmath,amssymb,amsfonts}
\usepackage{pifont}
\usepackage{xspace}
\usepackage{tabularx}
\usepackage[utf8]{inputenc}

\usepackage{threeparttable}
\usepackage{stfloats}
\usepackage{array,lipsum,afterpage,pdflscape}
\usepackage{footnotehyper}
\usepackage{breqn}
\usepackage{marvosym}
\usepackage{multirow}
\usepackage{makecell}
\usepackage[normalem]{ulem}
\usepackage{rotating}
\usepackage{tikz}
\usepackage{placeins}
\usepackage{cite}
\usepackage{scalerel}
\usepackage{tikz}
\usetikzlibrary{svg.path}

\definecolor{orcidlogocol}{HTML}{A6CE39}
\tikzset{
  orcidlogo/.pic={
    \fill[orcidlogocol] svg{M256,128c0,70.7-57.3,128-128,128C57.3,256,0,198.7,0,128C0,57.3,57.3,0,128,0C198.7,0,256,57.3,256,128z};
    \fill[white] svg{M86.3,186.2H70.9V79.1h15.4v48.4V186.2z}
                 svg{M108.9,79.1h41.6c39.6,0,57,28.3,57,53.6c0,27.5-21.5,53.6-56.8,53.6h-41.8V79.1z M124.3,172.4h24.5c34.9,0,42.9-26.5,42.9-39.7c0-21.5-13.7-39.7-43.7-39.7h-23.7V172.4z}
                 svg{M88.7,56.8c0,5.5-4.5,10.1-10.1,10.1c-5.6,0-10.1-4.6-10.1-10.1c0-5.6,4.5-10.1,10.1-10.1C84.2,46.7,88.7,51.3,88.7,56.8z};
  }
}

\newcommand\orcidicon[1]{\href{https://orcid.org/#1}{\mbox{\scalerel*{
\begin{tikzpicture}[yscale=-1,transform shape]
\pic{orcidlogo};
\end{tikzpicture}
}{|}}}}
\newcommand{\halfcircle}{\begin{picture}(6,6)
\put(3,3){\circle{6}}
\put(3,3){\circle*{3}}
\end{picture}}

\newtheorem{definition}{Definition}[section]

  {\endgroup} 
  
\ifCLASSOPTIONcompsoc
 \usepackage[caption=false,font=normalsize,labelfont=sf,textfont=sf]{subfig}
\else
 \usepackage[caption=false,font=footnotesize]{subfig}
\fi

\begin{document}
\title{Trust and Reputation in Data Sharing: A Survey}

\author{Wenbo~Wu,
        George~Konstantinidis
\thanks{Wenbo Wu and George Konstantinidis are with the School of Electronics and Computer Science, University of Southampton, Southampton, SO17 1BJ, UK (e-mail: \{wenbo.wu, g.konstantinidis\}@soton.ac.uk).}
}


\maketitle

\begin{abstract}
Data sharing is the fuel of the galloping artificial intelligence economy, providing diverse datasets for training robust models. Trust between data providers and data consumers is widely considered one of the most important factors for enabling data sharing initiatives. Concerns about data sensitivity, privacy breaches, and misuse contribute to reluctance in sharing data across various domains. In recent years, there has been a rise in technological and algorithmic solutions to measure, capture and manage trust, trustworthiness, and reputation in what we collectively refer to as \emph{Trust and Reputation Management Systems (TRMSs)}. Such approaches have been developed and applied to different domains of computer science, such as autonomous vehicles, or IoT networks, but there have not been dedicated approaches to data sharing and its unique characteristics. In this survey, we examine TRMSs from a data-sharing perspective, analyzing how they assess the trustworthiness of both data and entities across different environments. We develop novel taxonomies for system designs, trust evaluation framework, and evaluation metrics for both data and entity, and we systematically analyze the applicability of existing TRMSs in data sharing. Finally, we identify open challenges and propose future research directions to enhance the explainability, comprehensiveness, and accuracy of TRMSs in large-scale data-sharing ecosystems.
\end{abstract}

\begin{IEEEkeywords}
Trust and reputation, data sharing, data quality, decision making, regulatory compliance
\end{IEEEkeywords}

%

\section{Introduction}
\label{sec:introduction}
\IEEEPARstart{D}{ata} sharing is a key enabler for data ecosystems and markets, and a foundational asset for artificial intelligence (AI) technologies. The sharing of data among various users facilitates data aggregation, processing, and mining, while also underpinning innovative services and products that ultimately aid in economic growth and the improvement of quality of life~\cite{saura2021user,eu2020strategy}.
However, fostering broad user engagement in data sharing poses various challenges at the ethical, legal, cultural, financial, and technical levels~\cite{alter2018responsible,bonomi2020privacy}. Among these challenges, issues of trust between different roles (\textit{e.g.}, data providers and data consumers) are widely considered the primary impediment to effective data sharing~\cite{eu2020strategy,uk2022strategy}.

TRMSs~\cite{josang2007survey} act as an effective measure for addressing trust issues among interacting nodes in various networks. 
Traditional examples are found in e-commerce platforms (\textit{e.g.}, Amazon, eBay, Uber), where systems centered on user feedback allow customers to rate products or services. However, this reliance on human input is inherently subjective, and these systems often lack standardized evaluation methods.

Beyond commercial platforms, TRMSs are also integral to distributed networks such as the Internet of Things (IoT)~\cite{cao2016trust,jayasinghe2018machine,sagar2020trust,wei2020enhancing,liu2022semi}, multi-agent systems (MASs)~\cite{zikratov2016dynamic,muller2019trust,samuel2020trust,cheng2021general,Wu_2024}, and vehicular ad-hoc networks (VANETs)~\cite{lu2018bars,luo2019blockchain,kudva2021scalable,youssef2021distributed,liu2021behavior}. In these settings, they are deployed to assess node behaviors and continuously track their trustworthiness.
However, existing TRMSs face limitations due to their domain-specific and context-aware design, migrating them from one application to another is difficult. This challenge is prominent for data sharing ecosystems. Current TRMSs are incapable for this domain as they fail to account for the specific processes, roles, and evaluation metrics that characterize data sharing interactions~\cite{jussen2023data}.

In the context of data sharing, trust issues vary depending on the user's role and processes. For data providers, these concerns span the entire data lifecycle, stemming from risks to privacy and a lack of transparency in how their data is handled and processed~\cite{soria2016big,bonomi2020privacy,jussen2023data}.
First, there are security considerations regarding data intermediaries, such as whether data is protected by strong encryption in transit (e.g., using Transport Layer Security (TLS))~\cite{krawczyk2013security} and stored securely at rest. A more profound concern relates to data usage: whether the data consumer will strictly adhere to the agreements and purpose limitations~\cite{konstantinidis2021enabling}. This also involves ensuring that all processing activities comply with regulatory frameworks like the General Data Protection Regulation (GDPR)~\cite{abraham2019data,janssen2020data}.

The root of these concerns is the failure of current TRMSs to provide a mechanism for complete, trustworthy, and long-term tracking of data exchange activities within data markets and sharing platforms~\cite{huang2021toward,agarwal2019marketplace,koutsos2021agora,ma2024model}. Consequently, during data exchanges, users lack the context-aware trust and reputation references needed to make informed decisions about interacting with other nodes.

For data consumers, trust is a reciprocal concern revolving around the various dimensions of data quality (QoD)~\cite{pipino2002data,batini2009methodologies,evans2006scaling}.
The ingestion of low-quality data poses significant risks, as it can corrupt analytical outcomes, undermine the performance of machine learning models, and lead to flawed business intelligence~\cite{nguyen2025blockchain}. Consequently, foundational attributes such as data authenticity, integrity, and validity are key evaluation metrics. Consumers must be confident that the data originates from its claimed source, has not been altered without authorization, and conforms to the necessary standards for processing.

Beyond its basic form, the correctness and usability of data are determined by its accuracy, completeness, consistency, and reliability~\cite{pipino2002data,evans2006scaling}. For instance, a data consumer integrating datasets from multiple providers can find the process severely hampered by inconsistent formats or conflicting records, which may invalidate subsequent analysis~\cite{gupta2021data,santos2024can}. Data must also be fit for purpose, bringing dimensions like timeliness, relevance, and precision to the forefront; accurate but outdated data may be useless for real-time decisions, while irrelevant data can fail to meet analytical requirements~\cite{pipino2002data}. Finally, ensuring uniqueness and traceability is essential for avoiding skewed results from duplicate records and for auditing the data's provenance~\cite{budach2022effects}. These multifaceted quality concerns underscore the need for consumers to have robust trust models to assess data sources before engaging in an exchange.

Numerous pioneering studies offer solutions to evaluate QoD~\cite{pipino2002data,batini2009methodologies,evans2006scaling,li2022spatial}, audit legal compliance~\cite{garg2011policy,bichhawat2021automating,wang2022privguard}, and assess individual behaviors~\cite{cao2016trust,liu2021behavior,Wu_2024}. However, it is underexplored how these disparate solutions can be integrated into TRMSs. A comprehensive TRMS is needed that moves beyond one-off anomaly detection to incorporate these results into long-term records, using sound trust inference mechanisms to benefit future decision-making. 
Such TRMS directly integrates with core data engineering processes by translating complex assessments of data quality and entity behavior into quantifiable reputation scores, which in turn automate decisions—from weighting sources in data integration~\cite{tang2018reputation} and securing participant selection in federated learning~\cite{song2021reputation} to driving sophisticated incentive mechanisms in data markets~\cite{xiao2023resource}.

To systematically investigate the key elements and methodologies for building trust in data sharing, this survey develops novel taxonomies for TRMSs aimed at enhancing their explainability, comprehensiveness, and accuracy. Recognizing that existing trust models are often designed for domain-specific applications like IoT or VANETs, our approach is grounded in an analysis of the unique roles, processes, and evaluation metrics inherent to the data sharing ecosystem. The proposed taxonomies serve as a principled tool to deconstruct and evaluate TRMSs based on their core architectural and functional dimensions. By applying this framework, we identify essential characteristics required for the broader data-sharing context and highlight features that are currently underdeveloped, thereby laying a foundation for future research directions.
The main contributions of this survey are:
\begin{itemize}
    \item \textbf{TRMS Architecture for Data Sharing:} We introduce a novel TRMS architecture tailored to the data sharing ecosystem. Its primary purpose is to clarify the fundamental roles and functions of a TRMS within this specific context.
    \item \textbf{Novel Dual Taxonomies:} The survey introduces two new taxonomies. The first deconstructs the high-level system design of a TRMS. The second breaks down the core trust evaluation framework, distinguishing between the atomic trust signals and the computational models used for trust inference.

    \item \textbf{Data-Sharing Specific Evaluation Metrics:} We propose a comprehensive evaluation framework specifically for the data-sharing context. This framework moves beyond generic analysis to define distinct criteria for both data-centric quality and entity-centric behavior.
    
    \item \textbf{Systematic Analysis and Gap Identification:} Using the proposed taxonomies and metrics, the paper provides a systematic analysis of existing TRMSs across various domains. This review identifies a consistent gap in the literature: a predominant TRMSs focus on entity behavior while overlooking the quality of the shared data and the compliance of data consumers.
    
    \item \textbf{Forward-Looking Research Challenges:} This survey also identifies and details several overlooked open challenges that are critical for future research. 
\end{itemize}

\begin{table}[t]
\centering
\caption{Abbreviations in This Paper}
\label{tab:abbreviations}
\renewcommand{\arraystretch}{1.2} 
\begin{tabular}{ll}
\bottomrule
\textbf{Abbreviation} & \textbf{Description} \\
\midrule
AI & Artificial Intelligence \\
API & Application Programming Interface \\
DLT & Distributed Ledger Technology \\
DSA & Data Sharing Agreement \\
FL & Federated Learning \\
GDPR & General Data Protection Regulation \\
GNN & Graph Neural Network \\
IoT & Internet of Things \\
LLM & Large Language Model \\
MASs & Multi-agent Systems \\
PKI & Public Key Infrastructure \\
QoD & Quality of Data \\
RL & Reinforcement Learning \\
SLA & Service Level Agreement \\
SMPC & Secure Multi-Party Computation \\
TRMS & Trust and Reputation Management System \\
TTP & Trusted Third Party \\
VANETs & Vehicular Ad-hoc Networks \\
ZKP & Zero-Knowledge Proof \\
\toprule
\end{tabular}
\end{table}

The remainder of this paper is organized as follows: Section~\ref{sec:background} provides background knowledge on TRMS design in data sharing. Section~\ref{sec:TRMSs in data sharing} presents a novel TRMS architecture designed for data sharing ecosystem. Sections~\ref{sec:taxonomy} and~\ref{sec:trust evaluation framework} present system design taxonomy and trust evaluation framework, respectively. Section~\ref{sec:decision making} provides the decision-making strategies based on TRMSs. Section~\ref{sec:evaluation metrics} outlines the universal trust evaluation metrics for data sharing, while Section~\ref{sec:threats} summarizes the most prevalent threats and security issues in current TRMSs. In Section~\ref{sec:comparison}, we analyze and compare existing TRMSs using the proposed taxonomies. Building on these insights, we explore open challenges and opportunities for developing robust and reliable TRMSs to safeguard data sharing in Section~\ref{sec:challenges}. Finally, We conclude the paper in Section~\ref{sec:conclusion}. Table~\ref{tab:abbreviations} provides the most frequently used abbreviations in this paper.

\section{Background}
\label{sec:background}
This section establishes a foundation for analyzing TRMSs within data sharing ecosystems. First, we distinguish between the core concepts of trust, trustworthiness, and reputation. Second, we define the key roles, processes, and data usage rules that characterize data sharing. Finally, we review related work and delineate how our survey differs from prior research in its scope and contributions.

\subsection{Trust, trustworthiness, reputation}
\label{subsec:trust, trustworthiness reputation}
In both academic literature and practical systems, the concepts of trust, trustworthiness, and reputation are deeply interconnected yet distinct~\cite{richardson2003trust,josang2007survey,tennie2010reputation}. 

\begin{definition}[Trust]
    Trust is a subjective belief held by one entity (the trustor) about another (the trustee), representing the perceived probability that the trustee will perform a specific future action as expected, even without direct monitoring. 
\end{definition}
This belief is formed from both direct experience gained through prior interactions and indirect endorsements, such as recommendations from other known parties. More broadly, this subjective assessment is distinct from mere reliance, as trust involves an element of vulnerability and an assumption of the trustee’s good intentions, not just a calculation of their predictable behavior.

\begin{definition}[Trustworthiness]
    Trustworthiness is the inherent, objective quality of an entity that makes it deserving of trust; it is the property that trust and reputation management systems are ultimately designed to measure.
\end{definition}
This quality is widely understood to be multidimensional, commonly composed of different factors in different systems. An entity's overall trustworthiness is thus a function of these combined attributes.

\begin{definition}[Reputation]
    Reputation is the collective perception of an entity’s trustworthiness, derived from publicly accessible and aggregated information such as ratings, reviews, and referrals. 
\end{definition}
By summarizing the shared experiences of a community, it serves as a valuable and efficient signal for decision-making on platforms like Amazon. Beyond being an information source, reputation functions as a powerful social control mechanism that incentivizes cooperative behavior and deters malicious actions. It is typically built slowly over time through consistent performance but can be fragile and damaged quickly by negative events.

Based on the above definitions, the distinctions and connections between trust, trustworthiness, and reputation are presented. The discussion will now shift to the data sharing domain, where we will define the primary roles and processes involved in a data sharing ecosystem.

\subsection{Data sharing}
\label{subsec:data sharing}
Data sharing occurs across numerous mediums, including data marketplaces, model markets, and various distributed networks, each often operating with distinct standards and processes. Given this heterogeneity, a general definition is necessary to establish a common understanding that encompasses all such platforms. Accordingly, we adopt the definition from Jussen \textit{et al.}~\cite{jussen2023data}, where data sharing is defined as:
\begin{definition}[Data Sharing]
    Data sharing is the domain-independent process of giving third parties access to the data sets of others. These third parties may be other companies (usually not direct competitors), individuals, or public institutions. The shared data is often used to develop new applications and services. The expectation is to be compensated financially or through other benefits (\textit{e.g.}, receiving data) for providing the data. What the data may be used for and how it is made available is determined within the framework of the (legal) agreements between the data providers, data consumers, and other roles, depending on the use case.~\cite{jussen2023data}
\end{definition}

To provide a structured understanding of data sharing, we define four key roles commonly discussed in the literature (e.g., \cite{nguyen2025blockchain,jussen2023data,eichler2022data,schweihoff2023trust,castro2022protecting,schinke2023trustful}), each with distinct functionalities and responsibilities.

\begin{definition}[Data Provider]
    Data Provider is the individual or organization that originates or holds data and makes it available to others.
\end{definition}

\begin{definition}[Data Consumer]
    Data Consumer is the entity that accesses and utilizes data to generate insights, build applications, or inform decisions.
\end{definition}

\begin{definition}[Data Intermediary]
    Data Intermediary is a third-party platform or service that facilitates the exchange of data between data providers and data consumers. 
\end{definition}

\begin{definition}[Data Trustee]
    Data Trustee is an independent governance role responsible for the ethical and legal oversight of data within the ecosystem.
\end{definition}

Based on the defined roles, we can model the data sharing process as a sequence of three distinct phases, each characterized by different participating roles and objectives.

\noindent
\textbf{Preparation and Publication.} 
The Preparation and Publication phase constitutes the foundational activities undertaken exclusively within the data provider's domain to ensure a data asset is ready for exchange~\cite{eichler2022data}. This process begins with a strategic identification of the data's purpose and a rigorous governance and legal assessment, which includes establishing a lawful basis for sharing and conducting Data Protection Impact Assessments (DPIAs)~\cite{demetzou2019data} where necessary. A critical step involves data curation, encompassing cleansing, quality assurance, standardization into machine-readable formats, and the creation of detailed metadata to document the data's source, provenance, and schema~\cite{chen2013big,chapman2020dataset,paton2023dataset}. Concurrently, the provider drafts a legally binding Data Sharing Agreement (DSA)~\cite{swarup2006data} that codifies the permitted uses, restrictions, security obligations, and retention policies. The phase culminates in the technical publication of the curated and governed data asset, making it discoverable and securely accessible to potential consumers via mechanisms such as APIs or dedicated data marketplaces.

\noindent
\textbf{Data Sharing Transaction.}
The Sharing Transaction phase represents the critical interactive stage where the data provider and consumer engage directly or indirectly via data intermediary~\cite{castro2022protecting} to execute the data exchange. This phase is initiated by the data consumer, who discovers the data asset through a catalog or marketplace and evaluates its fitness-for-purpose using the provider's comprehensive metadata, often requesting a sample dataset for validation~\cite{schweihoff2023trust}. Following a successful evaluation, both parties negotiate and formally execute the DSA, establishing a legally binding framework for the interaction. The transaction concludes with the provider securely provisioning access—for instance, by issuing API keys or credentials—and the consumer subsequently ingesting the data into their own environment over secure, encrypted channels, marking the successful completion of the data transfer.

\noindent
\textbf{Utilization and Governance.}
Upon ingestion, the process enters the Utilization and Governance phase, where stewardship of the data asset transfers to the consumer's domain and is governed by a new set of obligations or a data trustee~\cite{schinke2023trustful}. Initially, the consumer integrates the acquired data with internal datasets and may perform further processing, ensuring all transformations adhere to the stipulations of the DSA. The primary objective is then realized through data utilization for value creation, such as training machine learning models, generating business intelligence, or informing strategic decisions. Critically, this phase requires the consumer to maintain ongoing compliance with the DSA, implement robust security controls to protect the data from unauthorized access or breaches, and manage the data's lifecycle, which includes its secure deletion upon the agreement's termination. This phase may also involve a feedback loop to the provider on data quality, thus contributing to the health of the broader data ecosystem.

Building upon the established data sharing processes, we now turn our attention to the data usage rules that are applied to safeguard the rights and interests of all users within the data sharing ecosystem.

\subsection{Data usage rules}
\label{subsec:data_protection}
\begin{table}
\centering
\caption{List of data protection regulations}
\label{tab:regu_poli}
\begingroup

\renewcommand{\arraystretch}{1.5} 
\begin{tabular}{p{0.08\linewidth} p{0.1\linewidth} p{0.63\linewidth}}
\toprule
\textbf{Name (Year)} & \textbf{Country Code} & \textbf{Description} \\ \midrule
PDPA (2012) \cite{pdpa_2012} & SG & Sets standards for personal data protection, including consent, notification, and retention policies. \\ \hline
GDPR (2018) \cite{gdpr_2018} & EU & Provides comprehensive rights to individuals regarding their personal data, including the right to access, correct, delete, and port data. Imposes strict consent and data handling requirements on organizations. \\ \hline
DPA (2018) \cite{dpa_2018} & UK & Aligns with GDPR, governs the processing of personal data, and provides specific legal rights for individuals in the UK. \\ \hline
CCPA (2020) \cite{ccpa_2020} & US & Empowers California residents to know what personal data is being collected, request deletion, and opt-out of the sale of their personal data. \\ 
\bottomrule
\end{tabular}
\endgroup
\end{table}

With growing awareness of data privacy and security, the implementation of robust data usage rules has become essential for fostering trust among data users. To facilitate the application of these rules within TRMSs, this paper categorizes them into three types based on their scope and nature: data protection regulations, data usage policies, and user preferences.

\noindent
\textbf{Data Protection Regulations.} Regulatory documents provide legal frameworks aimed at safeguarding the privacy and security of personal data. Table \ref{tab:regu_poli} lists four latest regulations enacted by various regional governments. These laws govern how data is collected, stored, processed, and shared, ensuring organizations adhere to strict compliance standards. 
For example, GDPR \cite{gdpr_2018} and CCPA \cite{ccpa_2020} aim to protect individual privacy by setting clear requirements for data handling and imposing penalties for non-compliance~\cite{custers2018comparison,merlec2021smart}. Despite these stringent penalties, numerous organizations have still been reported to experience data breaches in recent years\footnote{\url{https://www.enforcementtracker.com/}}.

\noindent
\textbf{Data Usage Policies.} Policies are organizational rules defining how data can be used and shared within specific contexts. These internal guidelines help clarify the scope and limitations of data usage to avoid violations and ensure ethical practices. Examples include enterprise data-sharing agreements and platform privacy policies \cite{benbernou2007privacy}, such as the terms of service provided by social media platforms. These policies not only establish boundaries for data usage but also promote operational transparency. By implementing such policies, organizations can align their practices with broader legal and ethical standards, fostering trust among stakeholders while mitigating the risks associated with improper data use.

\noindent
\textbf{User Preferences.} User consent mechanisms empower data providers to control how their personal data is used and shared. This is typically implemented through explicit consent forms or privacy settings that allow users to tailor their preferences~\cite{bonatti2021representing}. For instance, personalized user privacy preferences~\cite{holm2021control,chhetri2024enabling} are commonly used to give individuals greater control over their data. These mechanisms are particularly crucial in the era of smart devices, where extensive personal data is collected. However, traditional ``opt-in/opt-out'' systems often lack the granularity needed to address diverse privacy concerns fully. To address this gap, frameworks like the one proposed by Konstantinidis \textit{et al.}~\cite{konstantinidis2021enabling} enable customized consent within relational databases, while Park \textit{et al.}~\cite{park2023consent} developed a consent-based system for privacy-compliant data sharing. By respecting user preferences and providing granular control, user consent mechanisms establish trust in data sharing ecosystems.

\subsection{Distributed ledger technologies}
\label{subsec:dlt}
Distributed ledger technology (DLT) represents a decentralized system where multiple participants manage and validate transactions across network nodes. DLT can be categorized into three main types: blockchain, directed acyclic graph (DAG), and hybrid DLT, each offering distinct characteristics and applications~\cite{liu2020distributed}. With the rise of distributed systems such as IoT and VANET, DLT has gained popularity for its security features, including immutability, accountability, and traceability~\cite{liu2020distributed}. These attributes make DLT a valuable tool for TRMSs by securely documenting trust values associated with network entities~\cite{lahbib2024blockchain,hasan2022privacy}.

\noindent
\textbf{Blockchain.}
Blockchain, popularized by Bitcoin~\cite{nakamoto2008bitcoin}, operates on a decentralized architecture supported by cryptographic algorithms~\cite{zhang2019security} and consensus mechanisms~\cite{xiao2020survey}. Transactions are grouped into blocks, each cryptographically linked to its predecessor. Once confirmed, blocks become immutable and transparent to all participants. This structure ensures the integrity and security of data, as unauthorized modifications can be easily detected. Blockchain's robustness in maintaining trust has made it a cornerstone of DLT applications across multiple domains (\textit{e.g.}, VANETs~\cite{liu2021behavior}, IoT~\cite{liu2022semi}).

\noindent
\textbf{DAG-DLT.}
DAG-DLT offers an alternative to blockchain, with a structure where transactions directly validate previous ones without being grouped into blocks (\textit{e.g.}, IOTA~\cite{popov2018tangle}). In DAG-DLT, new transactions store and verify hashes of earlier transactions, streamlining validation and improving scalability by eliminating mining processes. This design enhances efficiency, particularly for high-throughput scenarios. However, the application of DAG-DLT in trust management remains an open research area, with potential for further exploration~\cite{fotia2023trust}.

\noindent
\textbf{Smart Contract.}
It was introduced by Szabo~\cite{szabo1997formalizing} and popularized by Ethereum~\cite{wood2014ethereum}, automate processes outlined in agreements without requiring a TTP. These self-executing programs, powered by blockchain, perform \texttt{Read-Modify-Write} operations and store results directly in the blockchain. Platforms like Quorum\footnote{\url{https://docs.goquorum.consensys.io/}} and Hyperledger Fabric\footnote{\url{https://hyperledger-fabric.readthedocs.io/en/release-2.5/}} have expanded smart contract applications.
Although smart contracts offer ACID (atomicity, consistency, isolation, durability) properties and eliminate intermediaries, challenges in security, privacy, and efficiency hinder their universal adoption~\cite{hu2021comprehensive,liu2024overview}.

\subsection{Related work}
\begin{table*}[t]
\caption{Comparison of our survey with existing surveys}
\label{tab:comparison_survey}
\centering
\begin{threeparttable}
\begingroup
\renewcommand{\arraystretch}{1.5} 
\begin{tabular}{llllllllll}
\bottomrule
\multirow{2}{*}{\textbf{Ref.}} &
\multirow{2}{*}{\textbf{Year}} & 
\multicolumn{1}{l|}{\multirow{2}{*}{\textbf{Topic}}} & 
\multicolumn{7}{c}{\textbf{Focuses}} \\ \cline{4-10} 
& 
& 
\multicolumn{1}{l|}{} & 
\ding{172} & \ding{173} & \ding{174} & \ding{175} & \ding{176} & \ding{177} & \ding{178}\\
\toprule

\cite{granatyr2015trust} & 2015 & 
Trust dimensions and interaction types of MASs in TRMSs. & 

\ding{109} & 
\ding{108} &
\ding{109} & 
\halfcircle&
\ding{108} & 
\ding{109} & 
\ding{109}\\ 
\hline

\cite{khan2020social} & 2020 & 
Quantitative and qualitative comparison of TRMSs in SIoT.  & 

\halfcircle & 
\ding{108} &
\ding{109} &
\ding{108} &
\ding{108} & 
\ding{109} &
\ding{109}
\\ 
\cline{1-10} 

\cite{sharma2020towards} & 2020 & 
Applications and trust management schemes of TRMSs in IoT. &

\ding{109} & 
\ding{109} & 
\ding{109} & 
\halfcircle &
\ding{108} & 
\ding{109} & 
\ding{109} \\ 
\cline{1-10} 

\cite{wei2022trust} & 
2022 & 
Analysis of IoT TRMSs based on the proposed hierarchical trust model. & 

\ding{109} & 
\ding{109} &
\halfcircle & 
\halfcircle &
\ding{108} & 
\ding{109} & 
\ding{109} \\ 
\cline{1-10}

\cite{wang2022heterogeneous} & 
2022 &
Trust models in heterogeneous networks. &

\ding{109} & 
\ding{108} & 
\halfcircle & 
\ding{108} & 
\ding{108} & 
\ding{109} &
\ding{109} \\ 
\hline

\cite{fotia2023trust} & 2023 & 
Convergence of edge computing and blockchain in TRMSs for IoT. & 

\ding{109} & 
\ding{109} &
\halfcircle & 
\ding{109} &
\ding{108} & 
\ding{109} & 
\ding{109} \\ 
\cline{1-10}
\hline

\cite{almarshoud2024security} & 
2024 & 
Decentralized trust management and privacy mechanisms in VANETs. &

\ding{109} & 
\ding{108} & 
\ding{109} & 
\halfcircle & 
\halfcircle &
\ding{109} &
\ding{109} \\ 
\hline

Ours & 
2025 &
TRMSs in data sharing and trust evaluation criteria &
\ding{108} &
\ding{108} &
\ding{108} &
\ding{108} &
\ding{108} &
\ding{108} &
\ding{108} 
\\
\toprule
\end{tabular}%
\endgroup
 \begin{tablenotes}\footnotesize
 \item   \ding{109}: not discussed;
         \ding{108}: discussed;
         \halfcircle: discussed implicitly without detailed classification and explanation.
 \item  \ding{172}: Characteristics of trust in data sharing, 
 \ding{173}: System design taxonomies for TRMSs, 
 \ding{174}: Trust evaluation framework, 
 \ding{175}: Security issues,  
 \ding{176}: Comparison and analysis of existing TRMSs,
 \ding{177}: Cross-domain applicability,
 \ding{178}: Open research challenges of TRMSs in data sharing.
 \end{tablenotes}
 \end{threeparttable}
\end{table*}
While many surveys on trust and reputation management exist, they are generally domain-specific due to the varying evaluation metrics required by different applications. This specificity results in a research gap, as existing work lacks a systematic focus on the processes, roles, and evaluation criteria particular to data sharing. This subsection aims to first review prior surveys and then delineating the unique contributions of our work, which are summarized for comparison in Tab.~\ref{tab:comparison_survey}.

In IoT networks, TRMSs facilitate informed decision-making by enabling information sharing in applications such as smart cities~\cite{cao2016trust}. Sharma \textit{et al.}~\cite{sharma2020towards} categorized TRMS approaches based on a hierarchical IoT architecture \cite{itu2012overview} and evaluated their applicability at the device and support layers. Their analysis highlights key research challenges for designing robust and scalable TRMSs for IoT data sharing, including heterogeneity, scalability, context-awareness, and resilience against attacks.


In a different approach, Wei \textit{et al.}~\cite{wei2022trust} identified a key gap in IoT trust management: that existing methods often overlook the complex interrelationships among three core components: trustworthiness calculation, object identification, and trust-based authorization. To bridge this gap, they proposed a three-layer model comprising incentive, computation, and data layers. 

Focusing specifically on edge-based IoT systems, the survey by Fotia \textit{et al.}~\cite{fotia2023trust} examined the convergence of edge computing and blockchain for designing secure and private TRMSs. The work provides a unique ``vertical'' analysis across the perception, network, and application layers, identifying key trust requirements like scalability and lightweightness.

Social IoT (SIoT) extends traditional IoT by enabling intelligent cooperation among devices through social relationships. Khan \textit{et al.}~\cite{khan2020social} explored TRMSs in this context, summarizing key research challenges. Their work combines a quantitative analysis grounded in fundamental trust concepts with a qualitative comparison of trust management processes, offering valuable insights for designing effective trust solutions in SIoT environments.

For trust management in MASs, Granatyr \textit{et al.}~\cite{granatyr2015trust} focused on the relationship between trust dimensions and interaction types (e.g., coalition, negotiation, recommendation), providing a unified framework for evaluating and designing TRMSs.

The survey by Almarshoud \textit{et al.}~\cite{almarshoud2024security} examined TRMSs in VANETs, with a focus on decentralized approaches, blockchain integration, and privacy-preserving mechanisms. It provides a taxonomy of trust-based services, identifies key requirements, and explores solutions such as SMPC and federated privacy to address challenges like trust distortion and the need for quantum-resistant cryptography.

Wang \textit{et al.}~\cite{wang2022heterogeneous} reviews TRMSs for heterogeneous networks (HetNets), highlighting the challenges posed by their unique characteristics, such as multi-domain involvement and openness. The study introduces the Quality of Trust (QoT) concept to evaluate trust models and systematically reviews their applications, strengths, and limitations. By addressing open issues like scalability, privacy, and cross-domain trust, the work provides a roadmap for building trustworthy HetNets, distinguishing itself from other domain-specific surveys by its broad scope.

While existing surveys offer valuable insights into TRMSs for various domains, they share a common limitation: a lack of specific focus on the data sharing context. In particular, they do not thoroughly explore the requirements for designing robust TRMSs capable of ensuring data quality and mitigating risks from unreliable participants in large-scale environments. Furthermore, most overlook the challenge of applying existing models to foster granular trust management that is explicitly tailored to data sharing.

\section{TRMSs in data sharing}
\label{sec:TRMSs in data sharing}

\begin{figure*}
    \centering
    \includegraphics[width=0.85\linewidth]{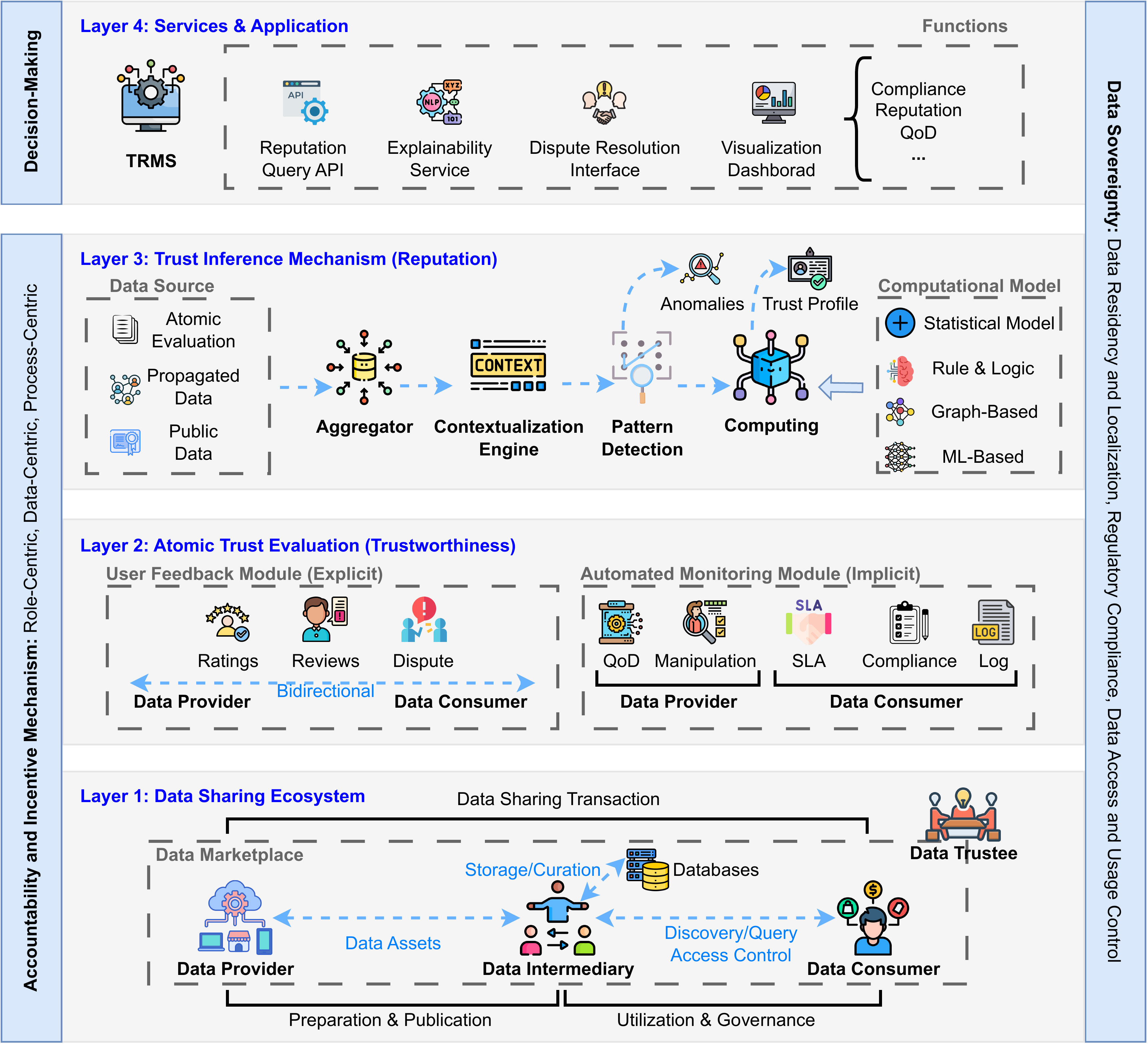}
    \caption{Vision Architecture of TRMS in Data Sharing}
    \label{fig:vision architecture}
\end{figure*}

Before presenting our taxonomies for TRMSs, we first propose a hierarchical architecture designed specifically for the data sharing ecosystem. This model serves to clarify the role and functions of a TRMS within this context. As illustrated in Fig.~\ref{fig:vision architecture}, the proposed architecture consists of four layers, each with distinct but interconnected functionalities. 
Collectively, these layers are designed to ensure data sovereignty by supporting data residency and localization, regulatory compliance, and fine-grained access and usage control. The first three layers provide the foundation for deploying accountability and incentive mechanisms, while the top layer offers an interface for data users to make informed decisions.

\noindent
\textbf{Layer 1: Data Sharing Ecosystem}
This layer represents the operational environment where data is exchanged. It consists of the core actors—Data Providers, Data Consumers, Data Intermediaries, and Data Trustee (e.g., marketplaces)—and the Data Assets themselves. The TRMS observes all interactions at this layer, such as data discovery, queries, transactions, and usage, to gather the raw evidence needed for trust evaluation.

\noindent
\textbf{Layer 2: Atomic Trust Evaluation (Trustworthiness)}
This layer is responsible for capturing raw, granular evidence of trustworthiness from the ecosystem. It acts as the sensory system for the TRMS which collect bidirectional, explicit and implicit trust signals.
Explicit signals are observed by the \texttt{Explicit Feedback Module}, which collects direct, human-driven inputs, including star ratings, textual reviews, and formal dispute filings from providers and consumers. 
Implicit signals are observed by the \texttt{Automated Monitoring Module}, which continuously monitors system interactions to generate objective, machine-driven evidence. It includes:
\begin{itemize}
    \item Data-Centric Monitor: Assesses the QoD by measuring metrics like accuracy, completeness, and timeliness.
    \item Compliance Monitor: Verifies adherence to rules and DSAs. This module could leverage LLMs to parse agreements and check against usage logs.
    \item Service Level Agreement (SLA) Monitor: Tracks operational metrics like API uptime and latency.
    \item Security Monitor: Logs security-relevant events, such as failed access attempts or potential data leakage.
\end{itemize}
The output of this layer is a continuous stream of atomic trust records, each detailing a single, evidence-backed event.

\noindent
\textbf{Layer 3: Trust Inference Mechanism (Reputation)}
This is the analytical core of the TRMS, where raw evidence is transformed into meaningful reputation scores.
The \texttt{Evidence Aggregator} collects atomic evaluation records, propagated information among network nodes, and publicly available information, then applies time-driven or event-driven strategies to combine them.
The \texttt{Contextualization Engine} provides adaptability. It takes into account the Application Domain (e.g., Healthcare, Finance), Entity Role, and specific goals to dynamically weight the importance of different metrics. For example, it would weigh Compliance more heavily in a healthcare context.
The \texttt{Pattern Detection Module} analyzes patterns in the evidence stream to detect and mitigate attacks as those discussed in Sec.~\ref{sec:threats}, like collusion or bad-mouthing, filtering or flagging malicious data before it can unfairly influence trust scores.
The \texttt{Computational Models} is where the actual reputation scores are calculated using a variety of models—statistical, graph-based, rule and logic-based, and machine learning-driven. These computational models are detailed in Sec.~\ref{subsec:trust inference mechanism}.

\begin{figure*}[t]
\centering
\includegraphics[width=0.9\textwidth]{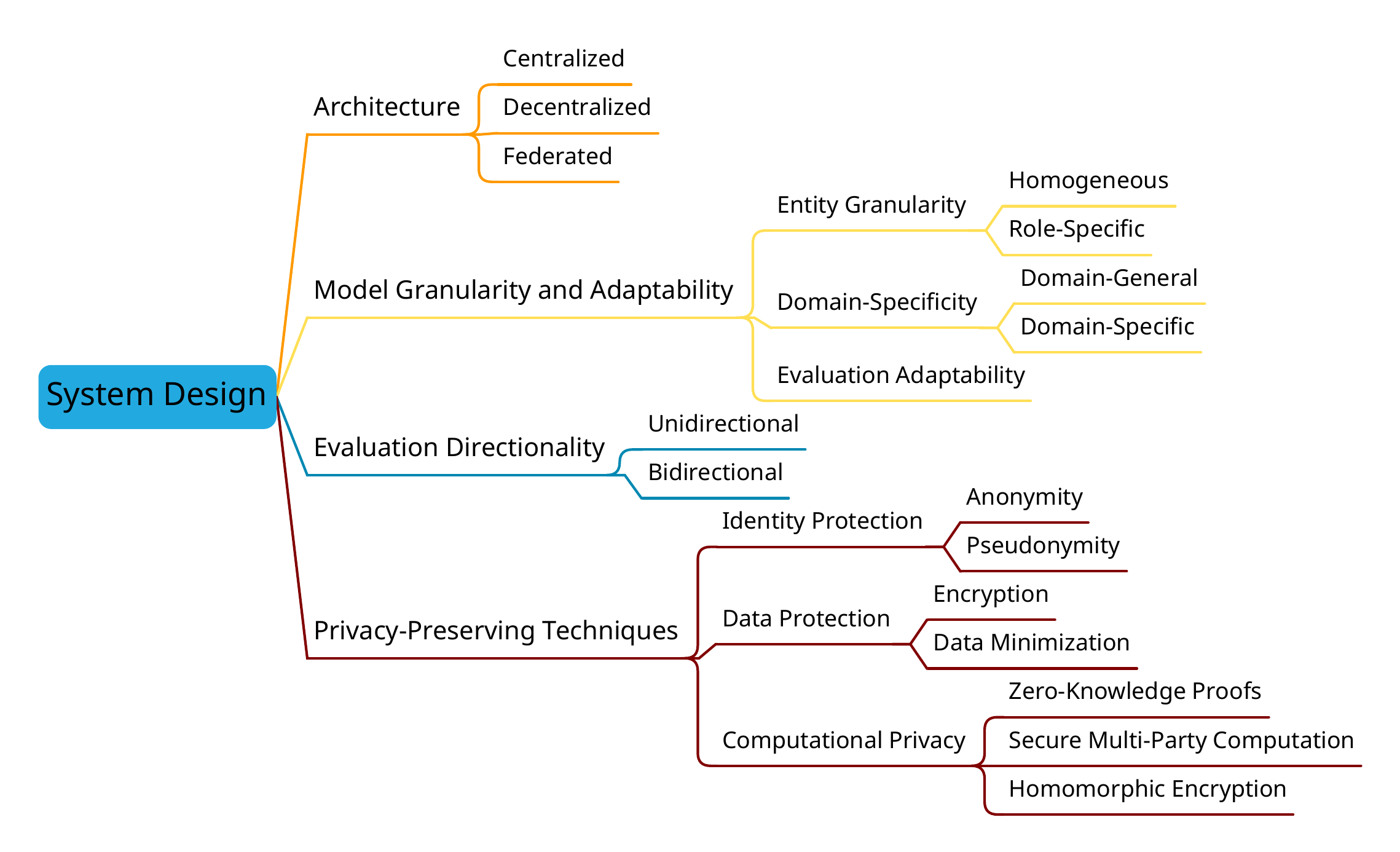}
\caption{System Design Taxonomy for TRMSs}
\label{fig:taxonomy of TRMS}
\end{figure*}

\noindent
\textbf{Layer 4: Services and Application Layer}
This top layer exposes the insights of the TRMS to users and external systems through a set of well-defined services.
The \texttt{Reputation Query API} allows authenticated users and applications to retrieve trust and reputation profiles for entities and data assets.
The \texttt{Explainability Service} makes the TRMS transparent. It uses techniques like SHAP/LIME or Natural Language Generation (from LLMs) to provide clear, human-readable justifications for any given trust score.
The \texttt{Dispute Resolution Interface} provides a user-facing portal where participants can manage disputes, backed by the governance framework and auditable logs.
The \texttt{Visualization Dashboard} provides a UI that presents comprehensive views of trust profiles, historical trends, compliance metrics, and data quality reports for stakeholders.

\section{System Design Taxonomy of TRMSs}
\label{sec:taxonomy}
Building upon the proposed architecture, which clarifies the role of a TRMS in the data sharing ecosystem, we now introduce a novel taxonomy to analyze the characteristics of such systems from a system design perspective (as illustrated in Fig.~\ref{fig:taxonomy of TRMS}). This taxonomy serves a dual purpose: on one hand, it allows for the systematic classification of existing literature and systems; on the other, it helps identify research gaps and opportunities, which are discussed further in Sec.~\ref{sec:challenges}. Our taxonomy is structured along four primary dimensions: architecture, model granularity and adaptability, evaluation directionality, and privacy-preserving techniques.

\subsection{Architecture}
The architecture of TRMSs is primarily divided into centralized, decentralized, and federated models, distinguished by the presence or absence of a Trusted Third Party (TTP) as well as its functionalities. These three models can be distinguished by their methods for trust inference, data storage, and their requirements for computing and communication resources.

\begin{figure*}
    \centering
    \includegraphics[width=\linewidth]{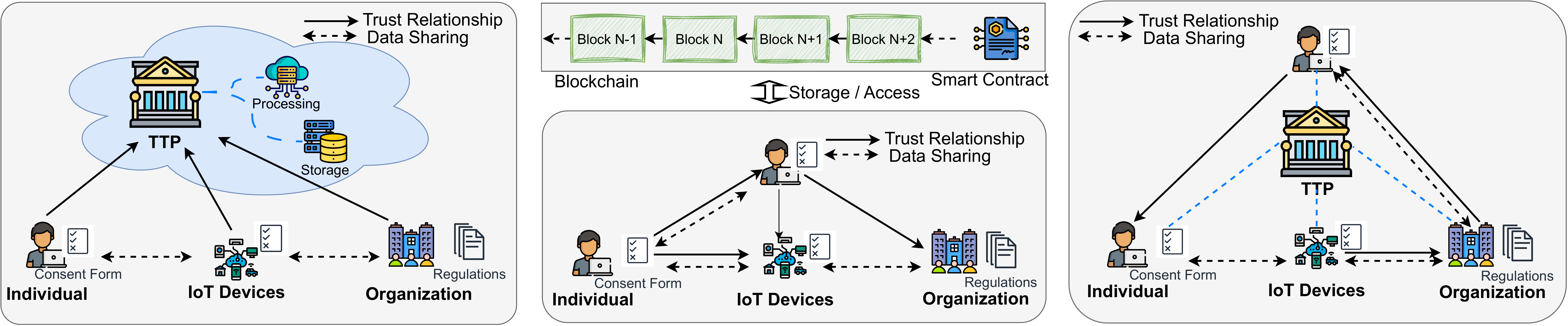}
    \caption{Centralized TRMS (left), Decentralized TRMS (middle), and Federated TRMS (right)}
    \label{fig:architecture}
\end{figure*}

\noindent
\textbf{Centralized model.} Centralized TRMSs operate under the governance of a TTP, which acts as the sole authority for all trust-related activities (as depicted in Fig.~\ref{fig:architecture}). Following each transaction, the TTP collects atomic trust evaluations, such as feedback and ratings, directly from the participating entities. It then applies trust inference mechanisms to aggregate these historical evaluations, thereby constructing a comprehensive trust profile for each entity~\cite{richardson2003trust}. This process typically yields a unified trust score, which is stored in a central database and made available to all network users as a reference. To maintain currency, the TTP dynamically updates these scores after every new transaction.

However, this centralized model presents several drawbacks. The resulting trust scores often provide limited information, lack explainability, and may be unsuitable for context-specific decisions. Furthermore, while the architecture simplifies evaluation by providing uniform metrics, it also creates a single point of failure, requires significant computational resources, suffers from scalability limitations, and introduces the risk of bias from the central authority.

\noindent
\textbf{Decentralized model.} Decentralized TRMSs eliminate the dependency on a TTP by distributing trust management responsibilities across all nodes in the network~\cite{fan2020decentralized}. This architecture is fundamentally enabled by technologies like Distributed Ledger Technology (DLT)~\cite{liu2020distributed}, which provides a transparent, immutable, and decentralized foundation for trust operations. In this paradigm, trustworthiness is evaluated by peers based on a combination of their direct interactions and information propagated from others  (as depicted in Fig.~\ref{fig:architecture}). Trust signals, including feedback and calculated scores, is stored securely and transparently across multiple nodes on a distributed ledger, often a blockchain~\cite{nguyen2025blockchain}. 

Updates to trust values are managed via consensus protocols, which require distributed computation and communication among participants to validate and record changes~\cite{alkharji2017authenticpeer,Wu_2024}. While this model offers enhanced system resilience, improved scalability, and reduced bias, it also introduces challenges such as consensus overhead, potential consistency issues, and a higher degree of complexity in defending against colluding malicious entities.

\noindent
\textbf{Federated Model.} The federated model combines elements of both centralized and decentralized approaches, seeking to balance their respective trade-offs~\cite{liu2022semi}. In this architecture, atomic trust evaluations are first performed locally, either by individual nodes or within distinct organizational or domain-specific clusters. Based on these local evaluations, each node maintains trust scores for its peers using its own computational model. These locally computed trust scores or vectors are then periodically transmitted to a semi-centralized aggregator. The aggregator's role is not to process every transaction but rather to synthesize these local trust assessments into a global reputation view, which is subsequently redistributed to all participants.

This model, conceptually similar to Federated Learning (FL)~\cite{mothukuri2021survey}, enhances scalability and reduces the communication overhead associated with purely centralized systems, while avoiding the consensus burden of fully decentralized ones. It is particularly well-suited for cross-domain ecosystems where participants wish to retain local control over their trust management processes while still benefiting from a shared, global reputation system.

\subsection{Evaluation granularity and adaptability}
The effectiveness and applicability of a TRMS are determined by its operational scope and the level of granularity at which it models the data sharing ecosystem. These characteristics can be analyzed across two fundamental dimensions: the level of detail at which participants are modeled (Entity Granularity), and the breadth of the system's applicability (Domain Specificity).

\noindent
\textbf{Entity Granularity.}
Entity granularity defines who is being evaluated and the level of specificity applied to different participants. TRMSs can either treat all entities as uniform actors or apply distinct rules and metrics based on their specific roles within the ecosystem.

Homogeneous models treat all participants as a uniform group, subject to the same set of trust evaluation rules. This approach is common in simpler peer-to-peer networks~\cite{tahta2015gentrust,naghizadeh2016c,alkharji2017authenticpeer,meng2020truetrust}, VANETs~\cite{lu2018bars,luo2019blockchain,kudva2021scalable,youssef2021distributed,liu2021behavior}, and IoT networks~\cite{cao2016trust,jayasinghe2018machine,sagar2020trust,wei2020enhancing,liu2022semi}, where entities perform similar functions.

Role-specific models apply distinct evaluation criteria tailored to the unique responsibilities of different roles, such as data provider, data consumer, or data intermediary~\cite{jussen2023data}. This allows for more precise and meaningful trust assessments, as the definition of trustworthy behavior differs significantly across roles. For example, a provider's trustworthiness may be a function of data quality~\cite{pipino2002data,batini2009methodologies,evans2006scaling}, while a consumer's is judged by their compliance with data usage policies~\cite{garg2011policy,bichhawat2021automating,wang2022privguard}. These models can be further distinguished by their flexibility in role assignment:
In single-role systems, entities are restricted to a fixed role, which streamlines evaluation. For example, in some healthcare contexts, individuals act solely as data providers, sharing records with institutions that are solely data consumers.
In multi-role systems, entities can perform multiple functions simultaneously. For instance, a vehicle in a VANET can act as both a data provider (sharing traffic data) and a data consumer (receiving route information), which enhances system adaptability but requires more complex trust mechanisms to evaluate performance in each distinct role~\cite{lu2018bars}.

\noindent
\textbf{Domain Specificity.}
Domain specificity defines where the trust model is intended to operate, distinguishing between systems designed for a single application context and those built for broad, cross-domain applicability.

Domain-specific models are tailored with metrics and logic optimized for a particular environment, such as e-commerce~\cite{bauer2020model} or social networks~\cite{lin2020guardian}. While these models can achieve high accuracy within their intended context, their specialized nature limits their portability and interoperability.

Domain-general models are designed to function across heterogeneous environments, a critical requirement for large-scale, multi-domain data marketplaces. Such systems must rely on more universal evaluation criteria and often require sophisticated mechanisms for trust transfer between different domains~\cite{awan2019holitrust}. For example, a privacy-preserving trust management architecture may use federated learning to build adaptable, task-specific trust models that can operate across diverse domains while maintaining a unified framework.

\noindent
\textbf{Evaluation Adaptability.}
Adaptability refers to a system's ability to adjust trust evaluations based on changing factors over time. It addresses the temporal nature of trust. Static trust models assume that trust is a fixed attribute, relying on historical data or one-time assessments~\cite{wilson1997certificates}. However, this fails to capture the fluid nature of trust in real-world interactions, which fluctuates based on time, context, and behavior \cite{wang2020survey,wang2022heterogeneous}.

Dynamic trust models address this limitation by continuously updating trust scores based on ongoing interactions. In data sharing ecosystems, for example, interactions between data providers and data consumers generate feedback that can dynamically refine trust assessments~\cite{wen2023dtrust}. These models also incorporate performance variations and external events, such as security breaches or policy changes, to maintain the relevance and accuracy of trust evaluations.

However, implementing dynamic adaptability introduces challenges, including computational complexity, vulnerability to malicious feedback, and the need for efficient consensus mechanisms in distributed systems. Overcoming these hurdles is essential to fully realize the benefits of adaptability in supporting secure and trustworthy data sharing ecosystems.

\subsection{Directionality}
The directionality of atomic trust evaluation (Layer 2 in Fig.~\ref{fig:vision architecture}), which dictates whether trustworthiness is assessed from a single perspective or through mutual evaluation among participants. This choice shapes the system's ability to provide a balanced and comprehensive view of the data sharing ecosystem.

\noindent \textbf{Unidirectional.}
In a unidirectional model, trust flows in a single direction, focusing on assessing the trustworthiness of one primary role. Such systems are common in traditional or constrained scenarios where the main perceived risk is associated with a specific type of entity, such as an unverified data provider. For instance, a TRMS might concentrate exclusively on evaluating a data provider's reliability by analyzing data authenticity and provenance, while largely ignoring the risks posed by a malicious data consumer~\cite{lu2018bars,liu2021behavior,inedjaren2021blockchain}. This inherent asymmetry, however, creates a significant vulnerability, as it leaves the system exposed to risks like unauthorized data usage or misinterpretation, which can undermine accountability and discourage broader participation.

\noindent \textbf{Bidirectional.}
Bidirectional models address the limitations of unidirectional approaches by enabling mutual evaluation, where each role can assess and be assessed by its counterparts. In a typical data provider-consumer arrangement, this allows consumers to rate data quality while enabling providers to verify consumer compliance with usage agreements~\cite{lu2018bars,liu2021behavior,inedjaren2021blockchain}. This creates a feedback loop that promotes continuous improvement and transparent accountability. The necessity for such models becomes even more pronounced in complex ecosystems where a single entity may hold multiple roles—for example, a hospital acting as both a data provider for research and a data consumer of patient data.

Despite the key advantages of greater transparency and balanced oversight, bidirectional models introduce significant operational complexity. They can be resource-intensive to maintain, particularly in decentralized environments where achieving real-time consensus for all roles is computationally demanding. Furthermore, their reliance on mutual ratings exposes them to greater risks of manipulation through collusive or Sybil attacks and necessitates more complex, context-aware analytics to manage entities holding multiple, distinct roles.

\subsection{Privacy-preserving techniques}
To ensure secure interactions and comply with data protection regulations, modern TRMSs must integrate robust privacy-preserving techniques designed to protect sensitive information throughout the trust evaluation lifecycle. For clarity, this survey categorizes these techniques based on their primary function into three main areas: identity protection, data protection, and computational privacy.

\noindent
\textbf{Identity Protection.}
Identity protection techniques focus on decoupling an entity's actions from its real-world identity, primarily through anonymity and pseudonymity~\cite{lu2018bars}. This is crucial for encouraging honest feedback, as participants may fear retaliation for negative ratings.

Anonymity aims to make an entity's actions unlikable to its identity. In a TRMS, this ensures that the source of a trust rating or piece of evidence cannot be traced back to the evaluating entity. For example, technologies like Tor or blockchain-based mixers can obscure the network origin of a transaction containing trust feedback, safeguarding user identities during trust evaluations~\cite{lahbib2024blockchain}.

Pseudonymity replaces real-world identifiers with consistent but unlinkable pseudonyms, often managed via a Public Key Infrastructure (PKI)~\cite{fromknecht2014decentralized}. In a typical deployment, as seen in the BARS framework for VANETs, a Certificate Authority (CA) issues pseudonymous public keys for privacy~\cite{lu2018bars}. This allows an entity to build a reputation tied to a pseudonym without revealing its real identity. For accountability, a separate, trusted Law Enforcement Authority (LEA) may hold the mappings between identities and keys, ensuring recourse in cases of malicious behavior.

\noindent
\textbf{Data Protection.} 
Data protection techniques safeguard the confidentiality and integrity of the trust data itself—both during storage and transit.

Encryption is a measure that renders data unreadable to unauthorized parties. In a TRMS, this involves using strong, standardized protocols like TLS~\cite{batini2009methodologies} for data in transit and robust algorithms like AES-256~\cite{abdullah2017advanced} for data at rest on servers or devices. Asymmetric encryption within a PKI~\cite{fromknecht2014decentralized} is also used to encrypt trust feedback directed at a specific recipient, ensuring only the intended party can decrypt it.

Data minimization is the principle of collecting, processing, and retaining only the data that is absolutely necessary for the trust computation. This directly supports compliance with regulations like GDPR's ``purpose limitation'' principle~\cite{gdpr_2018}. In a TRMS, instead of collecting detailed logs of an interaction, a system might only collect a binary outcome (e.g., success/failure) or a quantized rating (e.g., 1-5 stars)~\cite{kamvar2003eigentrust,aref2018hybrid}. This reduces the privacy risk, as less sensitive information is stored and available to be compromised.

\noindent
\textbf{Computational Privacy.}
Computational privacy techniques are advanced cryptographic methods that allow for the calculation of trust scores without centralizing or exposing the raw, sensitive inputs. This protects data even while it is being processed.

Zero-Knowledge Proofs (ZKPs) enable a prover to convince a verifier that a statement is true without revealing any information beyond the statement's validity. Modern ZKP systems like ZK-SNARKs or ZK-STARKs provide efficient methods for generating and verifying these proofs~\cite{aziz2025enhancing}.
In a data-sharing ecosystem, a participant could use a ZKP to prove their reputation score is above a certain threshold to gain access to a high-value dataset, without revealing their exact score~\cite{lu2008pseudo}. Similarly, an entity can prove that its rating was correctly included in an aggregated trust score update without revealing the rating itself, ensuring the integrity of the computation without sacrificing the privacy of its input.

SMPC allows multiple, non-trusting parties to jointly compute a function over their private inputs without any party revealing its input to others. Protocols often rely on techniques like secret sharing, where data is split into shares and distributed among computational nodes~\cite{zhao2019secure}.
SMPC is ideal for calculating a global reputation score in a decentralized ecosystem. For instance, several competing companies could use an SMPC protocol to compute the aggregate reputation of a shared software supplier~\cite{asharov2013fair}. Each company provides its private rating (e.g., number of service outages) as input. The SMPC protocol outputs the final aggregate score (e.g., average outages per month) to all participants without any company learning the specific input of another, thus preventing the leakage of sensitive business intelligence.

Homomorphic encryption enables mathematical operations to be performed directly on encrypted data (ciphertexts), generating an encrypted result that, when decrypted, matches the result of the operations as if they had been performed on the plaintext.
An aggregator in a TRMS can leverage an additively homomorphic scheme like the Paillier cryptosystem~\cite{wu2016reversible}. Participants encrypt their numerical ratings with the aggregator's public key and send the ciphertexts. The aggregator can then perform additions on these ciphertexts to compute the encrypted sum of all ratings and derive an encrypted average. The aggregator only ever decrypts the final aggregate score, meaning the individual ratings remain confidential even from the entity performing the calculation. While fully homomorphic encryption, which supports arbitrary computations, remains computationally intensive, it represents a powerful future tool for complex, privacy-preserving trust analytics~\cite{jency2025finquaxbot}.

\section{Trust evaluation framework}
\label{sec:trust evaluation framework}
Beyond the system-level design of a robust TRMS, the core component is the trust evaluation process itself. This section, therefore, we shift focus to trust evaluation, proposing a comprehensive trust evaluation framework. This framework is designed to encompass the entire evaluation lifecycle, from the initial atomic trust evaluation to the final trust inference mechanisms.

\begin{figure*}
\centering
  \includegraphics[width=\textwidth]{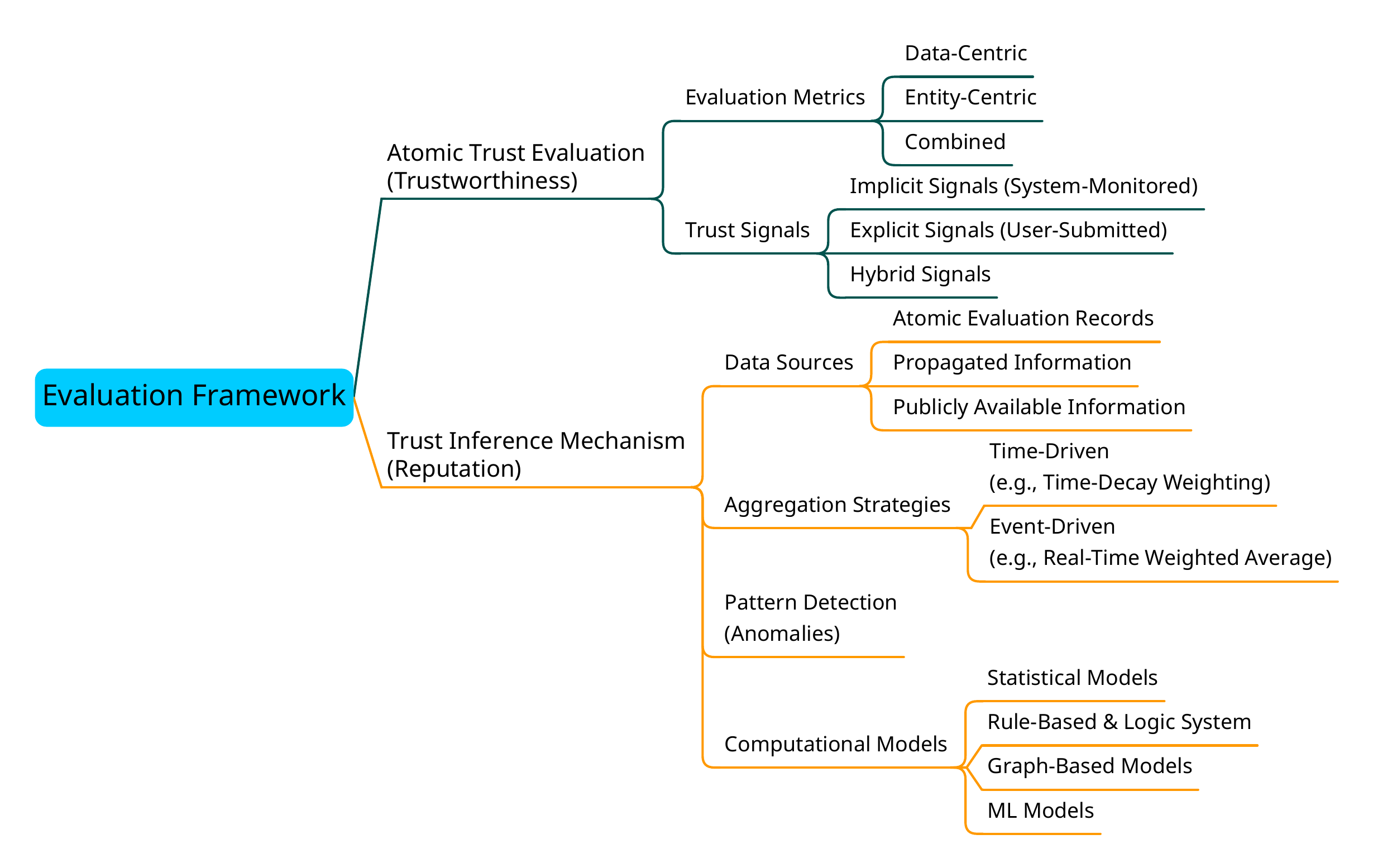}
\caption{Evaluation Framework of TRMSs}
\label{fig:taxonomy of trust evaluation} 
\end{figure*}

\subsection{Atomic trust evaluation}
Atomic trust evaluation forms the foundation of the trust assessment process. It refers to the initial, granular judgment of trustworthiness based on a single, direct interaction or a discrete piece of evidence. This atomic assessment serves as the fundamental building block for more complex reputation profiles. The process involves two key dimensions: the Evaluation Metrics that define what is being assessed, and the Trust Signals that provide the evidence for the assessment.

\noindent
\textbf{Evaluation Metrics.}
Evaluation metrics define the scope of the assessment, specifying whether the focus is on the data, the entity, or a combination of both. Generalizing from classifications originally developed for VANETs~\cite{ahmad2019comparative,yao2017using}, we identify three primary models for data-sharing ecosystems.

Data-centric models evaluate the trustworthiness of the shared data itself, emphasizing intrinsic properties such as accuracy, completeness, and authenticity. These models are critical in scenarios like machine learning, where data quality is a priority~\cite{gupta2021data}, and in data marketplaces, where data from unverified sources must be assessed~\cite{wang1996beyond}. A key principle of this approach is context dependency, as a dataset's trustworthiness often varies with its intended use~\cite{ebrahimi2022quantitative}. A notable example is the model by Raya \textit{et al.}, which evaluates data points independently of node reputation, making it resilient to malicious data injection in volatile ad-hoc networks~\cite{raya2008data}. Building on these principles, Sec.~\ref{subsec:data_centric metrics} of this survey proposes a unified trust evaluation framework to enable cross-domain QoD assessment for trustworthy data sharing.

Entity-centric models focus on evaluating the trustworthiness of participating entities (e.g., individuals, devices, organizations) by assessing their behavior, demonstrated properties, and relationships with others~\cite{conner2009trust,sateesh2020state}. These models are widely used in e-commerce and social platforms, where an entity's reputation is the primary signal for decision-making. For example, the model by Yao \textit{et al.} for VANETs evaluates the trustworthiness of vehicles to identify malicious nodes, using dynamic weighting to account for vehicle type and context~\cite{yao2017using}. In the context of large-scale data sharing, evaluating entity trustworthiness similarly requires a unified framework for comprehensive reputation construction, which we detail in Sec.~\ref{subsec:entity_centric metrics}.

Combined models integrate both data-centric and entity-centric approaches, simultaneously evaluating the trustworthiness of the data and the entities involved in sharing processes~\cite{mahmood2019hybrid}. By considering a broad range of factors, from data quality to entity behavior, these models provide a more nuanced and robust assessment suitable for complex ecosystems like healthcare, where both the integrity of the data and the credibility of its source are key factors~\cite{platt2018public}.

\noindent
\textbf{Trust Signals.}
Trust signals represent the evidence used to form an atomic assessment of trustworthiness. These signals serve as the primary inputs for the evaluation metrics and can be categorized based on their origin: implicit signals that are passively collected by the system, and explicit signals that are actively provided by users~\cite{jayasinghe2017data}.

Implicit Signals (System-Monitored) are derived from the algorithmic analysis of an entity's behavior against predefined, machine-readable rules~\cite{meng2018towards}. These rules, often based on standardized assessment criteria, allow trustworthiness to be updated iteratively and objectively after every interaction. By minimizing human bias and manual intervention, this approach provides real-time, consistent, and scalable trust evaluations, which are particularly effective in decentralized environments~\cite{wang2021c}. For instance, Wu \textit{et al.} proposed a framework for multi-agent systems where an agent's trustworthiness is assessed programmatically via a proof-of-location (PoL) verification mechanism, thereby enhancing the fairness and objectivity of the trust assessment~\cite{Wu_2024}.

Explicit Signals (User-Submitted) consist of subjective, user-provided feedback, such as reviews, ratings, or endorsements. This approach is prevalent on commercial platforms (e.g., Amazon, Uber) and promotes transparency and accountability through community validation~\cite{josang2007survey}. However, the primary challenge of explicit signals is their inherent subjectivity, as perceptions of trustworthiness can vary widely among users. Furthermore, these systems often lack explainability in how qualitative feedback is translated into a quantitative trust score, making the evaluation process opaque and vulnerable to manipulation.

Hybrid Signals combine both implicit (system-monitored) and explicit (user-submitted) signals to construct a more comprehensive and robust evaluation. By integrating automated behavioral analysis with subjective feedback, these models can enhance the credibility and accuracy of trust profiles, often incorporating human-in-the-loop (HITL) features~\cite{li2017human}. A common example involves pairing a rule-based system that detects malicious behavior with user feedback to validate and refine the overall trust evaluation~\cite{aref2018hybrid}. Despite their potential, hybrid models face challenges, including increased computational complexity and the difficulty of meaningfully quantifying non-numeric feedback, and thus remain an important area for future research.

\subsection{Trust inference mechanism}
\label{subsec:trust inference mechanism}
While atomic trust evaluation provides an initial assessment based on direct evidence, the trust inference mechanism (Layer 3 in Fig.~\ref{fig:vision architecture}) constructs a more comprehensive and resilient reputation profile. This is achieved by aggregating information from diverse sources to infer trustworthiness, particularly for entities with whom no direct interaction has occurred. As depicted in the Fig.~\ref{fig:taxonomy of trust evaluation}, this mechanism relies on four key components: the data sources that serve as inputs, the aggregation strategies that define how and when data is combined, the pattern detection systems that monitor for anomalies, and the computational models that perform the underlying calculations.

\noindent
\textbf{Data Sources.} 
The credibility of an inferred reputation score is contingent on the quality and diversity of its underlying data sources. A robust TRMS integrates evidence from multiple origins to build a holistic view of an entity's trustworthiness~\cite{guo2014merging}.

Atomic evaluation records are the most fundamental data source, consisting of the direct, first-hand trust assessments generated from the atomic trust evaluation phase. These records represent the ground truth of an entity's interactions within the system.

Propagated information leverages the principle of transitive trust, incorporating second-hand evidence from other entities in the network~\cite{guha2004propagation}. It answers the question: ``What do other trusted entities say about this entity?'' This is particularly powerful in decentralized systems where direct experience is sparse~\cite{guo2014merging}. For example, if Entity A trusts Entity B, and Entity B trusts Entity C, then Entity A can infer a degree of trust in Entity C.

Publicly available information incorporates external context from publicly accessible records to enrich the trust profile. This can include official certifications (e.g., ISO 27001 for security~\cite{calder2019information}), reports from TTPs, public security breach databases, or even sentiment analysis from news reports. This external validation provides an objective layer of verification that is difficult to manipulate from the ecosystem.

\noindent
\textbf{Aggregation Strategies.}
Aggregation strategies define the policy for when and how trust data is updated and combined into a reputation score—a critical design decision that balances system requirements like timeliness, stability, and computational efficiency~\cite{fan2020decentralized}. These strategies are generally categorized as either time-driven, operating on a fixed schedule, or event-driven, reacting instantly to new data.

Time-driven strategies update reputation scores on a periodic basis, making them predictable and computationally manageable in high-volume ecosystems. The most common method is the time-decay weighted average, where the influence of older evidence diminishes over time, allowing for reputation recovery and ensuring scores reflect recent behavior~\cite{jiang2020trust}. Another approach is fixed time-window averaging, which is simpler as it only considers a recent period of evaluations but can cause abrupt score changes and lacks long-term memory~\cite{zhou2015dynamic}. For high-security contexts, a more sophisticated method like scheduled Bayesian aggregation is used~\cite{wei2007trusted}. This statistically robust approach provides not just a score but also a confidence level by treating trust as a probability distribution, which is vital for distinguishing new entities from established ones.

In contrast, event-driven strategies update reputation scores in real-time following a new interaction or the receipt of new evidence, ensuring the score is always current. One direct method is the real-time cumulative average, where the score is recalculated after every rating~\cite{Wu_2024}; however, while highly responsive, this can be computationally expensive and volatile. A more stable alternative is exponential smoothing, which uses a smoothing factor to blend new ratings with the existing score, providing instant updates while preventing extreme fluctuations~\cite{wang2020dynamic}. A third, qualitatively different strategy is the event-triggered state change~\cite{williams2025event}. This approach allows a single critical report (e.g., a ``security failur'' flag) to bypass normal aggregation and trigger an immediate protective action, such as account suspension, serving as a vital safety layer against severe threats.

\noindent
\textbf{Pattern Detection.}
Beyond simple aggregation, a sophisticated inference mechanism must also actively monitor for anomalous patterns in trust data~\cite{li2019trust,qureshi2021anomaly}. This component acts as an early warning system, identifying deviations from an entity's established behavioral baseline or from the expected norms of the ecosystem. For example, a sudden, sharp decline in an entity's ratings could signal a compromised account or a significant drop in service quality. Likewise, detecting a large number of highly positive ratings from new, un-reputed accounts could indicate a ``ballot-stuffing'' attack. This is often implemented using anomaly detection algorithms, which fall under the umbrella of Machine Learning models.

\noindent
\textbf{Computational Models.}
Computational models are the algorithms used to implement the aggregation and pattern detection strategies, processing the various data sources to produce a quantitative reputation score. These models range from simple statistical methods to complex machine learning algorithms.

Statistical models: These are among the simplest and most widely used approaches. The sum and average methods are straightforward but susceptible to manipulation. A more robust approach is the weighted average, which assigns greater influence to ratings from more reputable sources, thereby mitigating the impact of biased or malicious feedback.

Rule-based and logic systems: These models leverage probabilistic and logical frameworks to manage uncertainty. Bayesian statistics provides a formal framework for updating trust beliefs as new evidence becomes available, using distributions like Beta or Dirichlet to model posterior probabilities~\cite{luo2019blockchain}. Dempster-Shafer Theory (DST) aggregates evidence from multiple sources while explicitly handling uncertainty and conflicting information~\cite{zhang2018novel}. Fuzzy logic allows for more granular trust levels (e.g., 'very low' to 'very high'), providing a context-sensitive evaluation~\cite{inedjaren2021blockchain}.

Graph-based models: These models are ideal for processing propagated information, representing the network of entities and their trust or endorsement relationships as a graph. Algorithms like PageRank~\cite{page1998pagerank} and EigenTrust~\cite{kamvar2003eigentrust} iteratively compute global reputation scores by aggregating trust values from an entity's neighbors, effectively modeling the flow of trust through the network.

Machine learning (ML) models: ML approaches utilize algorithms to learn complex patterns from interaction data and predict trustworthiness. Supervised learning (e.g., SVMs, Decision Trees) can classify entities, while unsupervised learning (e.g., k-means clustering and other anomaly detection algorithms) can identify patterns and outliers without labeled inputs. More advanced methods like Reinforcement Learning (RL) enable filtering of biased evaluation results~\cite{aref2018hybrid}, and Graph Neural Network (GNN) can improve the scalability and accuracy of trust aggregation in large-scale networks~\cite{lin2020guardian,wang2024trustguard}. However, ML models often face challenges in computational cost and explainability.

\section{Decision-Making Strategies with TRMSs}
\label{sec:decision making}
In open data sharing environments, effective decision-making helps with selecting trustworthy counterparts to maximize utility and ensure secure interactions. A TRMS provides the foundational intelligence for these decisions through the Layer 4 services (as depicted in Fig.~\ref{fig:vision architecture}), but this intelligence must be actioned through several distinct strategies. These can be broadly classified into: threshold-based and ranking strategies, AI-driven predictive strategies, automated policy-driven strategies, and explainability-driven human-in-the-loop strategies.

\noindent
\textbf{Threshold-Based and Ranking.}
The most direct methods are threshold-based and ranking strategies, where a consumer sets simple, objective rules based on the data retrieved from the TRMS~\cite{pitoura2022fairness}. This is facilitated by the \texttt{Reputation Query API}, which provides direct access to scores for QoD, Compliance, and overall Reputation. For example, a consumer might implement a threshold-based rule to automatically reject any data provider with a compliance score below 95\%. Alternatively, they could use a ranking strategy~\cite{kiefhaber2013ranking}: from a list of providers who meet a minimum QoD threshold, the system would select the one with the highest overall reputation score. These methods are computationally efficient and straightforward to implement for routine, low-risk decisions.

\noindent
\textbf{AI-Driven Predictive and Optimization.}
AI-driven predictive and optimization strategies represent a more sophisticated approach, leveraging machine learning to dynamically analyze complex and large-scale data from the TRMS~\cite{lin2020guardian,el2020machine}. Instead of relying on static thresholds, an AI agent can consume rich, time-series data from the \texttt{Reputation Query API} to predict future behavior, such as the likelihood of an SLA violation occurring in the next week. These strategies can also perform complex optimizations, for instance, by selecting a portfolio of data sources that maximizes predictive accuracy for a machine learning model while adhering to a predefined budget and risk tolerance based on the TRMS profiles. These adaptive methods are particularly effective in highly dynamic environments like IoT and e-commerce.

\noindent
\textbf{Automated Policy and Contract-Driven.}
Automated policy and contract-driven strategies focus on predefined mechanisms to enforce trust conditions without manual intervention. Game theory informs the design of the TRMS's underlying incentive and penalty systems, creating a framework where cooperative behavior is the most logical choice for all participants~\cite{barron2024game,wang2016game}. This is often implemented via smart contracts on blockchain platforms. For example, a Data Sharing Agreement can be encoded as a smart contract that automatically queries the TRMS. If the system's \texttt{Automated Monitoring Module} reports a critical DSA failure, the smart contract can instantly trigger a predefined penalty, such as slashing a portion of the provider's staked collateral~\cite{hasan2022privacy,zhaofeng2019blockchain,liu2021behavior}. This creates a transparent, self-enforcing trust environment that reduces reliance on intermediaries.

\noindent
\textbf{Explainability-Driven.}
Finally, explainability-driven, human-in-the-loop strategies are essential for high-stakes, novel, or complex decisions where full automation is undesirable~\cite{inedjaren2021blockchain,zhaofeng2019blockchain}. For these scenarios, a human decision-maker leverages the full suite of Layer 4 services. For instance, before entering a major partnership, a manager might use the \texttt{Visualization Dashboard} to conduct a deep analysis of a provider's historical performance. If they notice an anomaly, they can use the \texttt{Explainability Service} to receive a natural-language summary explaining why a trust score dipped three months ago. This detailed, justifiable insight provides the context needed for a confident, high-stakes decision. This same service is critical during formal disputes, where the \texttt{Dispute Resolution Interface} can present an auditable, explainable history to all parties, ensuring a fair and transparent process.

\section{Trust evaluation metrics in data sharing}
\label{sec:evaluation metrics}
In Sec.~\ref{sec:background}, we discuss data sharing from the perspectives of its constituent roles and processes, highlighting the ecosystem's diverse characteristics, inherent challenges, and the potential for TRMSs to enhance interactions. These elements serve as the foundation for addressing trust-related issues when designing a TRMS. Building on this foundation, we propose a set of trust evaluation metrics for data sharing, with a specific focus on assessing both data attributes and entity behaviors. These metrics are designed to improve the reliability of trust evaluations and provide a robust framework for safeguarding the entire data sharing process.

\subsection{Data-centric metrics}
\label{subsec:data_centric metrics}
\begin{figure}
    \centering
    \includegraphics[width=0.9\linewidth]{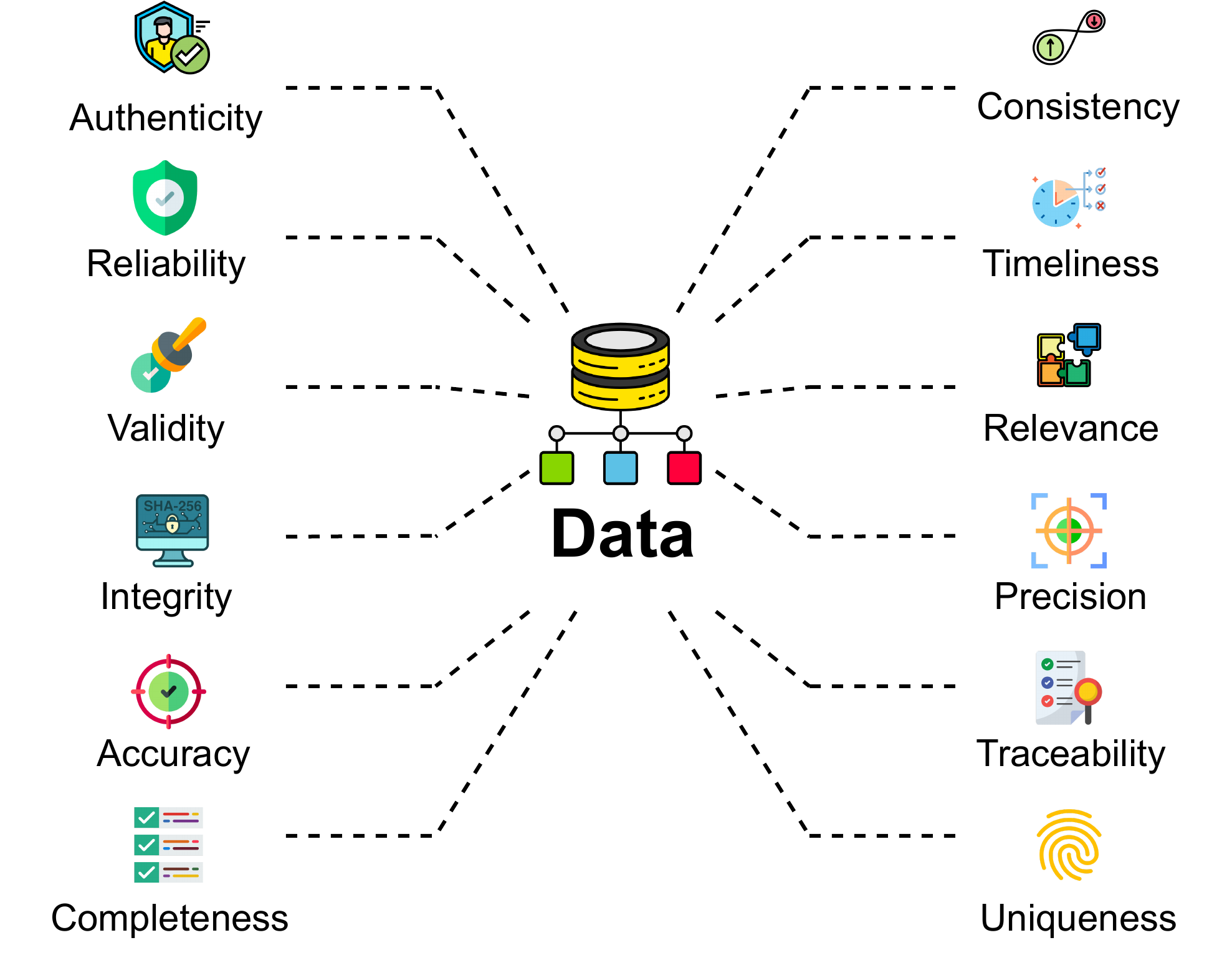}
    \caption{Trust evaluation dimensions on data}
    \label{fig:criteria_data}
\end{figure}

The trustworthiness of a data provider is usually closely related to data it provides, and the QoD~\cite{batini2009methodologies,gudivada2017data,budach2022effects} that can be measured from multiple dimensions (as depicted in Fig.~\ref{fig:criteria_data}).

\textit{Authenticity} of data refers to the genuineness of data, ensuring that it is the original data produced by its claimed source without unauthorized alterations~\cite{rahman2021lossless,chen2024event}. In the context of data quality, authenticity is used for validating the source, authorship, and origin of data, ensuring it has not been tampered with or misrepresented. It is closely linked to data integrity, which ensures that data remains complete, consistent, and unaltered from its original form throughout its lifecycle. Authenticity involves verifying the data's provenance, ensuring non-repudiation (\textit{i.e.}, ensuring that the data source cannot deny its authorship), employing mechanisms such as digital signatures or encryption.

\textit{Reliability} refers to the degree to which data consistently reflects stable and dependable information over its lifecycle. It emphasizes the trustworthiness of data, ensuring that it reliably represents reality without unexpected variations~\cite{meeker2022statistical}. Reliable data maintains its quality across different contexts and uses, providing a steady foundation for confident decision-making. By prioritizing reliability, organizations can trust that the data will consistently support robust and insightful analyses, even as it undergoes regular use and handling within various systems.

\textit{Validity} refers to the extent to which data meets the requirements of specified formats, structures, and standards, thereby ensuring its appropriateness for intended use~\cite{skaaning2018different}. 
For instance, numerical data should be in the correct numeric format, date fields must follow standardized date conventions, and categorical data must fit predefined classifications. Validity also requires conformance to standards such as ISO or other industry-specific guidelines, which ensure compatibility and consistency across systems. Meeting these standards enhances its effectiveness in high-stakes decision-making, promoting trusted and seamless integration across diverse applications.

\textit{Integrity} refers to the preservation of data’s original state and the ability to monitor any changes made to it over time. Unlike dimensions that focus on the correctness of the data or the confidence in its source, integrity is specifically concerned with whether the data remains unaltered from its original form, free from unauthorized or unintended modifications \cite{harley2021information}. One component of integrity is auditability, which involves maintaining detailed records of any transformations or updates applied to the data. 

\textit{Accuracy} represents the extent to which data correctly reflects the real-world entities or events it is designed to capture, serving as a measure of closeness between data values and their actual, ``ground truth'' counterparts. Accuracy can be further categorized into feature accuracy and target accuracy: feature accuracy assesses how well individual data attributes match reality, while target accuracy focuses on the correct representation of overarching data targets or objectives~\cite{budach2022effects}. For any data type, accuracy is maintained by ensuring each attribute falls within a defined acceptable range; deviations outside these bounds indicate inaccuracies. Errors, distortions, or mismatches in data attributes compromise accuracy, reducing the data’s fidelity to real-world conditions.

\textit{Completeness} refers to the degree to which all necessary data is present and available \cite{batini2009methodologies}, ensuring that there are no missing elements that could compromise analysis or decision-making. This dimension is evaluated by checking that each data item includes all required attributes and that each attribute has corresponding values~\cite{batini2009methodologies}. High completeness is essential for robust model predictions and accurate assessments, as missing data can undermine the reliability of outcomes and reduce confidence in conclusions drawn from the data.

\textit{Consistency} ensures uniformity and coherence within and across datasets, reducing the risk of contradictory or misleading information. A consistent dataset upholds standardized formats, units, and representations, enabling seamless integration and accurate analysis~\cite{chakraborty2015network,gudivada2017data}. For example, discrepancies in units of measurement or data formats across datasets can create confusion and errors during interpretation. Similarly, internal inconsistencies, such as conflicting values or duplicate records may result in flawed conclusions.

\textit{Timeliness} (or freshness) refers to the extent to which data is current and accessible at the precise moment it is needed, a highly relevant factor in time-sensitive domains such as weather forecasting and traffic prediction. As data ages, its ability to accurately represent real-world events diminishes, reducing its relevance and potentially leading to outdated insights~\cite{gudivada2017data}. Evaluating timeliness involves assessing how well the data reflects present conditions and whether it is available within a timeframe that preserves its value for decision-making. In scenarios where timing is essential, timeliness directly influences the data’s utility, determining its effectiveness in supporting real-time, informed decisions.

\textit{Relevance} refers to the extent to which data is applicable and valuable for the specific purpose it is intended to serve. High relevance ensures that the data aligns with the goals of the analysis or application, providing meaningful insights that drive informed decisions~\cite{doku2019towards,streufert1973effects}. When data is relevant, it directly addresses the needs of the analysis, enhancing the accuracy and effectiveness of the outcomes. In data sharing, ensuring relevance becomes even more important, as the data must be tailored to meet the requirements of various stakeholders and applications. If the data lacks relevance, it may lead to wasted resources and misguided conclusions, ultimately diminishing its utility.

\textit{Precision} refers to the level of detail or granularity present in the data, which can determine its suitability for the intended analysis. When data is sufficiently detailed, it allows for a more accurate and nuanced understanding of the subject matter, supporting in-depth analysis and robust decision-making. The required level of precision depends on the specific needs of the decision-making process~\cite{thapa2021precision}; for example, highly granular data may be essential for precision medicine~\cite{blasimme2018data}, while broader data may suffice for more general analyses~\cite{cao2016trust}.

\textit{Traceability} refers to the capacity to track data throughout its lifecycle, documenting each step from its point of origin through various stages of processing and transformation. This dimension ensures that every change or movement in the data’s journey is recorded, providing a transparent, verifiable history~\cite{gudivada2017data,huang2020blockchain}. Traceability allows stakeholders to see how data has evolved, ensuring that all processes it has undergone are clear and accountable~\cite{yu2021blockchain}. By establishing a comprehensive audit trail, traceability supports confident use of the data, as users can confirm its handling and modifications over time, which is especially valuable in contexts where understanding the data’s lineage is essential for validating its reliability and context in decision-making.

\textit{Uniqueness} represents the extent to which a dataset is free from unnecessary duplication, ensuring each entity or event is captured only once. This dimension preserves the dataset’s clarity and operational efficiency, as duplicate records or redundant data points can introduce distortions and lead to inefficient analysis~\cite{gudivada2017data}. A lack of uniqueness can cause misleading conclusions, inflated metrics, and resource wastage by skewing results with repeated information~\cite{akbar2024enhanced}. Maintaining uniqueness eliminates these risks, offering a streamlined and dependable dataset that supports accurate analysis and effective decision-making.

\subsection{Entity-centric metrics}
\label{subsec:entity_centric metrics}

\begin{figure}[t]
    \centering
    \includegraphics[width=0.9\linewidth]{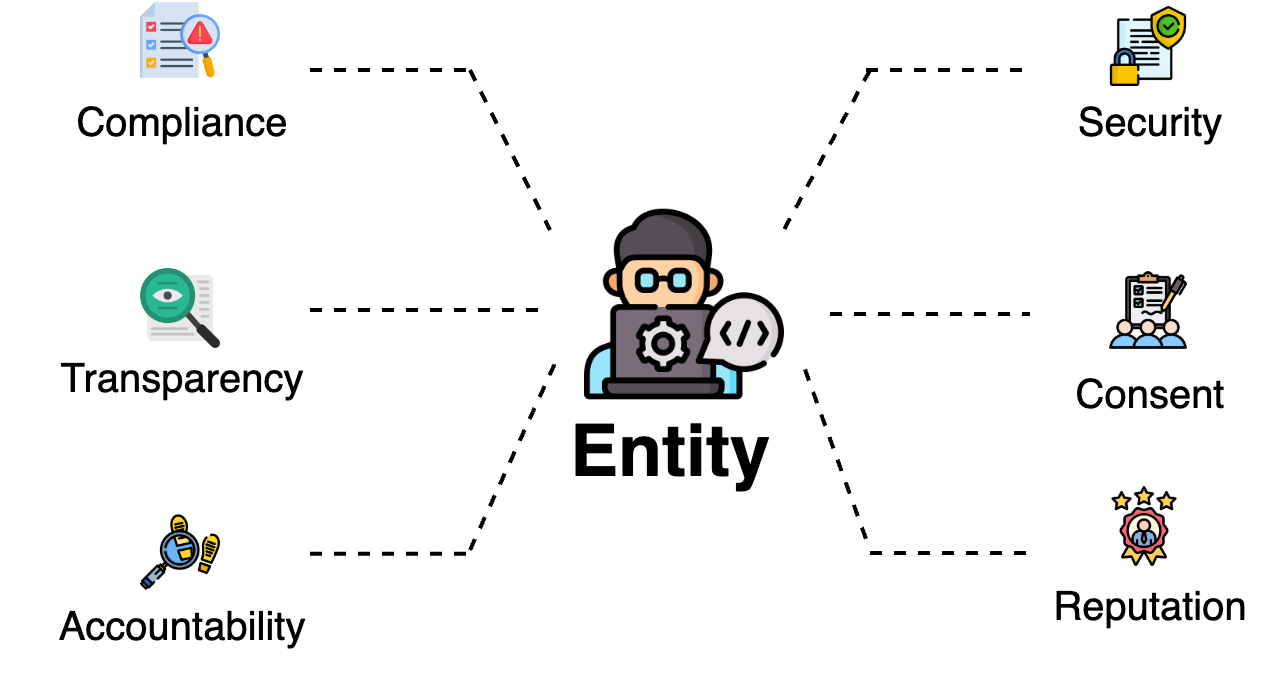}
    \caption{Trust evaluation dimensions on entity}
    \label{fig:criteria_entity}
\end{figure}

The trustworthiness of a data consumer is typically tied to their data processing activities and behaviors~\cite{bertino2017big,alneyadi2016survey,wang2018not}, as illustrated in Fig.~\ref{fig:criteria_entity}.
To address the objectives discussed in Sec.~\ref{sec:background}, trust evaluation metrics tailored to entities are focusing on dimensions like accountability, compliance, reputation, and reliability. 

\textit{Compliance} refers to the adherence to established data sharing rules and regulations. This includes ensuring that data sharing practices align with the intended purpose and intent for which the data was collected and shared~\cite{chhetri2022data,yalamati2024data}. Compliance can maintain trust in data sharing environments, as it ensures that DPs and DCs follow legal, ethical, and operational guidelines, thereby protecting data integrity and respecting user consent~\cite{holm2021control,chhetri2024enabling,chhetri2022data}. Adherence to these rules not only safeguards the privacy and rights of individuals but also ensures that the shared data is used appropriately and in line with agreed-upon terms~\cite{cao2016trust}.

\textit{Transparency} refers to the degree to which the DCs openly communicates how data is used, shared, and stored, establishing a foundation of trust through clear, accountable, and user-centered practices. Transparency requires data consumers to openly disclose how they collect data, including specifying sources, acquisition methods, and the intended purpose of collection~\cite{jin2022peekaboo}. Purpose specification is another key element, as data consumers should clearly communicate the reasons for data processing, ensuring these purposes align with user expectations and comply with relevant legal frameworks. 

\textit{Accountability} reflects an organization’s commitment to responsible data management and protection. A robust governance structure involving clear accountability channels and the designation of dedicated roles such as a data protection officer (DPO) or privacy officer who oversees compliance and data protection initiatives~\cite{park2022open}. Additionally, internal audits or third-party assessments are key components of accountability, as regular evaluations of data processing practices help identify potential risks, ensure adherence to policies, and reinforce transparency. Together, these mechanisms promote a culture of responsibility and continuous improvement, demonstrating the organization’s dedication to safeguarding data integrity and building trust with stakeholders.

\textit{Security} practices are safeguarding data from unauthorized access and breaches, ensuring robust protection measures are in place throughout the data's lifecycle. Security measures require the implementation of robust technical and organizational protections, such as encryption, access controls, and anonymization techniques, to ensure data confidentiality and integrity~\cite{saura2021user}.

\textit{Consent} is focusing on the data consumer's respect for users' autonomy over their personal information. Effective consent mechanisms ensure that users are fully informed and able to give explicit, granular consent for specific data uses, while also allowing them to withdraw that consent easily and at any time~\cite{konstantinidis2020need}. Furthermore, data subject rights are critical, as users are supposed to be empowered to exercise their rights under relevant legal frameworks, including access to their data, the ability to request corrections, deletion, or portability of their data. Upholding these rights can enhance user trust, as it ensures that individuals retain control over their personal information and that the data consumer adheres to legal and ethical standards.

\textit{Reputation} of data provider refers to the degree to which data is considered trustworthy and originates from a reputable source. It encompasses the reliability and regard of the data source, as well as whether the data has been verified or validated by trusted authorities~\cite{ma2021trusted}. Additionally, reputation is shaped by the historical behavior of entities, including their cumulative performance in providing accurate and reliable data over time~\cite{xie2021secure}. References and endorsements from other reputable entities further bolster the DP's reputation, enhancing the overall trustworthiness of the data in data-sharing environments~\cite{li2011exploiting}.
The reputation of a data consumer refers to the degree to which it is regarded based on its historical behaviors and endorsements from others. A DC can enhance its reputation by strictly adhering to the data processing agreements established with the DPs, demonstrating reliability and trustworthiness in handling shared data.

\section{Threats and security issues}
\label{sec:threats}
Beyond its core architecture, the development of a secure TRMS for data sharing also accounts for a wide array of potential threats. Many existing TRMSs lack full resilience against sophisticated attacks—a limitation noted in several studies~\cite{lahbib2024blockchain,josang2007survey}. Accordingly, this section provides an overview of key threats that target TRMSs. Our scope is threat identification; for a detailed analysis of countermeasures, readers are directed to surveys such as~\cite{alneyadi2016survey,bampatsikos2021solving,wang2022heterogeneous}.

\textit{Sybil attack}~\cite{zhou2011p2dap} refers to a malicious user generates numerous fake identities, known as Sybils, to exert undue influence over the system. This strategy can lead to the artificial boosting of certain entities' reputations or the unwarranted tarnishing of competitors' reputations. The feasibility of a Sybil attack is contingent on the cost associated with acquiring new identities; thus, increasing the cost to create new identities can diminish the risk of such attacks. The most effective approach involves linking the identity within the system to a real-world, unique identifier of the corresponding entity, thereby making the creation of fake identities more challenging~\cite{dennis2015rep}.

\textit{Privacy} concerns are also existing in trust and reputation management systems~\cite{lu2018bars}, just as they are in other research domains~\cite{baruh2017online}. These systems typically assess the trustworthiness of network entities by collecting and analyzing user data to determine behavior and assign reputation scores. While PKI~\cite{spies2017public} based systems offer basic security services in many distributed systems, safeguarding the privacy of each identity within a trust and reputation framework poses challenges. This situation highlights issues regarding user privacy, data protection, and the risk of sensitive information being misused.

Newly joined users or entities in a system often face the challenge of having insufficient data available for an accurate assessment of their trustworthiness, a situation known as the ``\textit{cold start} trust score'' or ``initial trust score problem''~\cite{bampatsikos2021solving}. This can leave them vulnerable to unfair judgements or manipulation. To tackle this issue, most trust management approaches resort to setting an arbitrary or neutral initial trust score, some may overlook this challenge entirely. Assigning arbitrary trust scores to newly connected devices can disrupt system operations, either by mistakenly granting high trust scores to malicious devices or by unfairly treating trustworthy devices as potential threats, thereby impairing system performance.

In scenarios where a user with a poor reputation opts to abandon their identity and create a new one, they effectively reset their reputation score for \textit{whitewashing}, evading the repercussions of their previous actions. This attack proves efficient not only due to the low cost of entering the network but also because the system ranks an entity with a zero reputation score higher than a user with negative ratings~\cite{anceaume2013privacy}. This provides an incentive for users to abandon their accounts once their reputation becomes negative, thereby circumventing the consequences of their actions by starting afresh with a neutral reputation score. A potent countermeasure, similar to the most effective strategy against Sybil attacks, involves linking the system identity of an entity to a unique, real-world identifier~\cite{lu2018bars}. This approach ensures that each entity can only register a single identity in the system, making them traceable and accountable for their actions.

\textit{Ballot stuffing} (\textit{i.e.}, self-promotion) involves users artificially inflating their reputation via coordinated actions or multiple accounts under their control. This can deceive other users about the genuine trustworthiness of the entity being promoted. It can be mitigated by limiting token creation, which curtails feedback opportunities and could lead to financial losses from missing genuine transactions~\cite{lahbib2024blockchain}. Beyond financial repercussions, such actions could also discourage engagement with truly trustworthy individuals. While participants might buy extra tokens which will incur additional costs. Beyond financial repercussions, such actions could also discourage engagement with truly trustworthy individuals. Distributed ledger technology (DLT) has emerged as an effective strategy to prevent the impact of these attacks, ensuring a more secure and reliable environment~\cite{liu2020distributed}.

The \textit{bad-mouthing} (\textit{i.e.}, slandering) attack involves spreading false information about the data quality of a provider or the behavior of a consumer to reduce their reputation, making it a direct and simple form of attack. Unlike ballot stuffing, which aims to artificially boost an entity's reputation, bad-mouthing seeks to damage competitors' or others' reputations in the system with unfounded or exaggerated negative feedback. This can skew the perceived trustworthiness within the system and negatively affect innocent parties. Bankovic \textit{et al.}~\cite{bankovic2011detecting} proposed a method to detect and isolate such attackers in IoT networks using outlier detection with the self-organizing maps (SOM) algorithm. Their simulation results demonstrated that this approach could effectively identify malicious entities within the system.

In many distributed networks, \textit{on-off attack} (\textit{i.e.}, strategic manipulation) refers to entities that behave honestly for most of the time to build trust and subsequently engage in malicious activities can be difficult to detect promptly~\cite{caminha2018smart}. TRMSs often interpret occasional misconduct as a temporary error, allowing such attackers to continually disrupt normal system operations undetected. These strategic manipulators can thus maintain their presence within the system for an extended duration. To conceptualize this type of attack, when a malicious node alternates between \(n\) good behaviors and \(m\) bad behaviors, it is termed an \(n\)G-\(m\)B On-off attack~\cite{chae2014trust}. 

\textit{Collusion} represents a prevalent attack in reputation systems, where multiple users conspire to manipulate reputation scores, either by mutually enhancing each other's reputation (collusive boosting) or by collectively targeting another entity (collusive slandering)~\cite{chahal2020trust}. The orchestrated nature of collusion makes it particularly challenging to detect and prevent. 
Shen \textit{et al.}~\cite{shen2015enhancing} proposed collusion detection methods to enhance the resilience of TRMSs against malicious rating manipulation in P2P networks, e-commerce, and social platforms. It introduces a basic detection method, an optimized approach to reduce computational overhead, and pre-processing techniques (random-cut and k-means clustering) to efficiently identify colluding nodes.

\begin{table*}
\centering
\caption{Comparison of Trust Evaluation Metrics in Distributed Autonomous Systems}
\label{tab:metric_DAS}
\begin{threeparttable}

\begingroup

\renewcommand{\arraystretch}{1.5} 
\begin{tabular}{l|lllllllllllllllllll}
\bottomrule
\multirow{2}{*}{\textbf{Domain}} &
\multirow{2}{*}{\textbf{Reference}} &
\multicolumn{12}{|c}{\textbf{Data}\tnote{I}} &  
\multicolumn{6}{|c}{\textbf{Entity}\tnote{II}}  
\\
\cline{3-14} \cline{15-20}  
 &  & \multicolumn{1}{|l}{\textbf{Au}} & \textbf{Ra} & \textbf{Va} & \textbf{In} & \textbf{Ac} & \textbf{Cp} & \textbf{Cs} & \textbf{Ti} & \textbf{Rl} & \textbf{Pr} & \textbf{Tr} & \textbf{Un} & \multicolumn{1}{|l}{\textbf{Cl}} & \textbf{Tp} & \textbf{Ao} & \textbf{Se} & \textbf{Co} & \textbf{Re} \\ 
\toprule
  
\multirow{6}{*}{\textbf{VANETs}} 
                        & \cite{lu2018bars} 2018 & \ding{108} & \ding{109} & \ding{108} & \ding{108} & \ding{108} & \ding{109} & \ding{109} & \ding{109} & \ding{109} & \ding{109} & \ding{108} & \ding{109} & \ding{109} & \ding{109} & \ding{108} & \ding{108} & \ding{109} & \ding{108} \\
                        & \cite{luo2019blockchain} 2019 & \ding{108} & \ding{109} & \ding{109} & \ding{109} & \ding{109} & \ding{109} & \ding{109} & \ding{109} & \ding{109} & \ding{109} & \ding{108} & \ding{109} & \ding{109} & \ding{109} & \ding{108} & \ding{108} & \ding{109} & \ding{108} \\
                        & \cite{kudva2021scalable} 2021 & \ding{109} & \ding{109} & \ding{109} & \ding{108} & \ding{109} & \ding{109} & \ding{109} & \ding{109} & \ding{109} & \ding{109} & \ding{108} & \ding{109} & \ding{109} & \ding{108} & \ding{108} & \ding{108} & \ding{109} & \ding{108} \\
                        & \cite{youssef2021distributed} 2021 & \ding{108} & \ding{108} & \ding{109} & \ding{108} & \ding{109} & \ding{109} & \ding{109} & \ding{109} & \ding{109} & \ding{109} & \ding{108} & \ding{109} & \ding{109} & \ding{109} & \ding{108} & \ding{108} & \ding{109} & \ding{108} \\
                        & \cite{liu2021behavior} 2021 & \ding{108} & \ding{109} & \ding{109} & \ding{109} & \ding{109} & \ding{109} & \ding{108} & \ding{109} & \ding{109} & \ding{109} & \ding{108} & \ding{109} & \ding{109} & \ding{109} & \ding{108} & \ding{108} & \ding{109} & \ding{108} \\  
                        \hline
\multirow{5}{*}{\textbf{IoT}} 
                        & \cite{cao2016trust} 2016 & \ding{109} & \ding{109} & \ding{109} & \ding{109} & \ding{109} & \ding{109} & \ding{109} & \ding{109} & \ding{109} & \ding{109} & \ding{109} & \ding{109} & \ding{109} & \ding{108} & \ding{108} & \ding{109} & \ding{108} & \ding{109} \\
                        & \cite{jayasinghe2018machine} 2018 & \ding{109} & \ding{109} & \ding{109} & \ding{109} & \ding{109} & \ding{109} & \ding{109} & \ding{109} & \ding{109} & \ding{109} & \ding{109} & \ding{109} & \ding{109} & \ding{109} & \ding{109} & \ding{109} & \ding{109} & \ding{108} \\
                        & \cite{sagar2020trust} 2020 & \ding{109} & \ding{109} & \ding{109} & \ding{109} & \ding{109} & \ding{109} & \ding{109} & \ding{109} & \ding{109} & \ding{109} & \ding{109} & \ding{109} & \ding{109} & \ding{109} & \ding{109} & \ding{109} & \ding{109} & \ding{108} \\
                        & \cite{wei2020enhancing} 2020 & \ding{109} & \ding{109} & \ding{109} & \ding{109} & \ding{109} & \ding{109} & \ding{109} & \ding{109} & \ding{109} & \ding{109} & \ding{109} & \ding{109} & \ding{109} & \ding{109} & \ding{108} & \ding{108} & \ding{109} & \ding{108} \\
                        & \cite{liu2022semi} 2023 & \ding{109} & \ding{109} & \ding{109} & \ding{109} & \ding{109} & \ding{109} & \ding{109} & \ding{109} & \ding{109} & \ding{109} & \ding{109} & \ding{109} & \ding{109} & \ding{109} & \ding{108} & \ding{109} & \ding{109} & \ding{108} \\
                        \hline  
\multirow{5}{*}{\textbf{MASs}} 
                        & \cite{zikratov2016dynamic} 2016 & \ding{109} & \ding{109} & \ding{108} & \ding{108} & \ding{109} & \ding{109} & \ding{109} & \ding{109} & \ding{109} & \ding{109} & \ding{109} & \ding{109} & \ding{109} & \ding{109} & \ding{109} & \ding{109} & \ding{109} & \ding{108} \\
                        & \cite{muller2019trust} 2019 & \ding{108} & \ding{109} & \ding{109} & \ding{108} & \ding{108} & \ding{109} & \ding{109} & \ding{109} & \ding{109} & \ding{109} & \ding{108} & \ding{109} & \ding{109} & \ding{109} & \ding{109} & \ding{109} & \ding{109} & \ding{108} \\ 
                        & \cite{samuel2020trust} 2020 & \ding{109} & \ding{109} & \ding{109} & \ding{109} & \ding{109} & \ding{109} & \ding{109} & \ding{109} & \ding{109} & \ding{109} & \ding{109} & \ding{109} & \ding{109} & \ding{109} & \ding{109} & \ding{108} & \ding{109} & \ding{108} \\ 
                        & \cite{cheng2021general} 2021 & \ding{109} & \ding{109} & \ding{109} & \ding{109} & \ding{108} & \ding{109} & \ding{109} & \ding{109} & \ding{109} & \ding{109} & \ding{109} & \ding{109} & \ding{109} & \ding{109} & \ding{109} & \ding{109} & \ding{109} & \ding{109} \\ 
                        & \cite{Wu_2024} 2024 & \ding{108} & \ding{109} & \ding{108} & \ding{108} & \ding{109} & \ding{109} & \ding{109} & \ding{109} & \ding{109} & \ding{109} & \ding{108} & \ding{109} & \ding{109} & \ding{109} & \ding{109} & \ding{108} & \ding{109} & \ding{108} \\  
                        \hline
\multirow{4}{*}{\textbf{P2P}} 
                        & \cite{tahta2015gentrust} 2015 & \ding{108} & \ding{109} & \ding{109} & \ding{108} & \ding{108} & \ding{109} & \ding{109} & \ding{109} & \ding{109} & \ding{109} & \ding{109} & \ding{109} & \ding{109} & \ding{109} & \ding{109} & \ding{109} & \ding{109} & \ding{108} \\ 
                        & \cite{naghizadeh2016c} 2016 & \ding{109} & \ding{109} & \ding{109} & \ding{109} & \ding{109} & \ding{109} & \ding{109} & \ding{109} & \ding{109} & \ding{109} & \ding{109} & \ding{109} & \ding{109} & \ding{109} & \ding{109} & \ding{109} & \ding{109} & \ding{108} \\
                        & \cite{alkharji2017authenticpeer} 2017 & \ding{108} & \ding{109} & \ding{108} & \ding{108} & \ding{109} & \ding{109} & \ding{109} & \ding{109} & \ding{109} & \ding{109} & \ding{109} & \ding{109} & \ding{109} & \ding{109} & \ding{109} & \ding{108} & \ding{109} & \ding{108} \\ 
                        & \cite{meng2020truetrust} 2020 & \ding{108} & \ding{109} & \ding{109} & \ding{109} & \ding{109} & \ding{109} & \ding{109} & \ding{109} & \ding{109} & \ding{109} & \ding{109} & \ding{109} & \ding{109} & \ding{109} & \ding{109} & \ding{109} & \ding{109} & \ding{108} \\
                        \hline
\toprule
\end{tabular}%
\endgroup
 \begin{tablenotes}
 \item  \ding{109}: not supported;
        \ding{108}: supported.
 \item I: \textbf{Au:} Authenticity, \textbf{Ra:} Reliability, \textbf{Va:} Validity,  \textbf{In:} Integrity, \textbf{Ac:} Accuracy, \textbf{Cp:} Completeness, \textbf{Cs:} Consistency, \textbf{Ti:} Timeliness \textbf{Rl:} Relevance, \textbf{Pr:} Precision, \textbf{Tr:} Traceability, \textbf{Un:} Uniqueness.
 \item II: \textbf{Cl:} Compliance, \textbf{Tp:} Transparency, \textbf{Ao:} Accountability, \textbf{Se:} Security, \textbf{Co:} Consent, \textbf{Re:} Reputation.
 \end{tablenotes}
 \end{threeparttable}
\end{table*}

\begin{table*}[t]
\caption{System Design of TRMSs in Distributed Autonomous Systems}
\label{tab:design_DAS}
\resizebox{\textwidth}{!}{%
\renewcommand{\arraystretch}{1.5} 
\begin{tabular}{l|lllp{0.07\linewidth}llccc}
\bottomrule
\multirow{3}{*}{\textbf{Domain}} &
  \multicolumn{1}{l}{\multirow{3}{*}{\textbf{Reference}}} &
  \multirow{3}{*}{\textbf{Architecture}} &
  \multirow{3}{*}{\textbf{\begin{tabular}[c]{@{}l@{}}Evaluation\\ Granularity\end{tabular}}} &
  \multirow{3}{*}{\textbf{\begin{tabular}[c]{@{}l@{}}Domain\\ Specificity\end{tabular}}} &
  \multirow{3}{*}{\textbf{\begin{tabular}[c]{@{}l@{}}Evaluation\\ Adaptability\end{tabular}}} &
  \multicolumn{1}{l|}{\multirow{3}{*}{\textbf{\begin{tabular}[c]{@{}l@{}}Evaluation\\ Directionality\end{tabular}}}} &
  \multicolumn{3}{c}{\textbf{Privacy-Preserving Techniques}} \\ \cline{8-10} 
 &
  \multicolumn{1}{l}{} &
   &
   &
   &
   &
  \multicolumn{1}{l|}{} &
  \textbf{\begin{tabular}[c]{@{}l@{}}Identity\\ Protection\end{tabular}} &
  \textbf{\begin{tabular}[c]{@{}l@{}}Data\\ Protection\end{tabular}} &
  \textbf{\begin{tabular}[c]{@{}l@{}}Computational\\ Privacy\end{tabular}} \\ \toprule
\multirow{5}{*}{\textbf{VANETs}} 
 & \cite{lu2018bars} 2018 & Centralized & Homogeneous & Specific & Dynamic & Unidirectional & \ding{51} & \ding{51} & \ding{55} \\
 & \cite{luo2019blockchain} 2019& Decentralized & Role-Specific & Specific & Dynamic & Bidirectional & \ding{51} & \ding{51} & \ding{55} \\
 & \cite{kudva2021scalable} 2021& Decentralized & Homogeneous & Specific & Dynamic & Unidirectional & \ding{51} & \ding{51} & \ding{55} \\
 & \cite{youssef2021distributed} 2021 & Decentralized & Homogeneous & Specific & Dynamic & Unidirectional & \ding{55} & \ding{51} & \ding{55} \\
 & \cite{liu2021behavior} 2021 & Decentralized & Homogeneous & Specific & Dynamic & Bidirectional & \ding{51} & \ding{51} & \ding{55} \\
\hline

\multirow{5}{*}{\textbf{IoT}} 
 & \cite{cao2016trust} 2016    & Centralized & Role-Specific & Specific & N/A & N/A & \ding{55} & \ding{51} & \ding{55} \\
 & \cite{jayasinghe2018machine} 2018 & Centralized & Homogeneous & General & Dynamic & Unidirectional & \ding{55} & \ding{55} & \ding{55} \\
 & \cite{sagar2020trust} 2020  & Centralized & Homogeneous & Specific & Dynamic & Unidirectional & \ding{55} & \ding{55} & \ding{55} \\
 & \cite{wei2020enhancing} 2020& Centralized & Homogeneous & Specific & Dynamic & Unidirectional & \ding{55} & \ding{51} & \ding{55} \\
 & \cite{liu2022semi} 2022     & Federated & Homogeneous & General & Dynamic & Unidirectional & \ding{55} & \ding{51} & \ding{55}\\
\hline

\multirow{5}{*}{\textbf{MASs}}
 & \cite{zikratov2016dynamic} 2016 & Decentralized & Homogeneous & Specific & Dynamic & Unidirectional & \ding{55} & \ding{55} & \ding{55} \\
 & \cite{muller2019trust} 2019 & Decentralized & Homogeneous & Specific & Dynamic & Unidirectional & \ding{51} & \ding{51} & \ding{55}\\
 & \cite{samuel2020trust} 2020 & Decentralized & Homogeneous & Specific & Dynamic & Unidirectional & \ding{55} & \ding{51} & \ding{55} \\
 & \cite{cheng2021general} 2021 & Centralized & Homogeneous & General & Dynamic & Unidirectional & \ding{55} & \ding{55} & \ding{55} \\
 & \cite{Wu_2024} 2024  & Decentralized & Homogeneous & Specific & Dynamic & Unidirectional & \ding{51} & \ding{55} & \ding{55} \\
\hline

\multirow{4}{*}{\textbf{P2P}}
 & \cite{tahta2015gentrust} 2015 & Decentralized & Homogeneous & Specific & Dynamic & Unidirectional & \ding{55} & \ding{55} & \ding{55}\\
 & \cite{naghizadeh2016c} 2016 & Decentralized & Homogeneous & Specific & Dynamic & Unidirectional & \ding{55} & \ding{55} & \ding{55}\\
 & \cite{alkharji2017authenticpeer} 2017 & Decentralized & Homogeneous & Specific & Dynamic & Unidirectional & \ding{55} & \ding{55} & \ding{55} \\
 & \cite{meng2020truetrust} 2020 & Decentralized & Homogeneous & Specific & Dynamic & Unidirectional & \ding{55} & \ding{55} & \ding{55}\\ \toprule
\end{tabular}
}
\end{table*}
\begin{table*}[]
\caption{Trust Evaluation Framework of TRMSs in Distributed Autonomous Systems}
\label{tab:evaluation_DAS}
\centering
\resizebox{\textwidth}{!}{%
\renewcommand{\arraystretch}{1.5} 
\begin{tabular}{l|llllllllll}
\bottomrule
\multicolumn{1}{c|}{\multirow{3}{*}{\textbf{Domain}}} &
  \multicolumn{1}{c}{\multirow{3}{*}{\textbf{Reference}}} &
  \multicolumn{1}{c}{\multirow{3}{*}{\textbf{\begin{tabular}[c]{@{}c@{}}Evaluation\\ Metrics\end{tabular}}}} &
  \multicolumn{1}{c|}{\multirow{3}{*}{\textbf{\begin{tabular}[c]{@{}c@{}}Trust\\ Signals\end{tabular}}}} &
  \multicolumn{3}{c|}{\textbf{Data Sources}} &
  \multicolumn{2}{c|}{\textbf{\begin{tabular}[c]{@{}c@{}}Aggregation\\ Strategies\end{tabular}}} &
  \multicolumn{1}{l}{\multirow{3}{*}{\textbf{\begin{tabular}[l]{@{}l@{}}Pattern\\ Detection\end{tabular}}}} &
  \multicolumn{1}{c}{\multirow{3}{*}{\textbf{\begin{tabular}[c]{@{}l@{}}Computational\\ Models\end{tabular}}}} \\ \cline{5-9}
\multicolumn{1}{c|}{} &
  \multicolumn{1}{c}{} &
  \multicolumn{1}{c}{} &
  \multicolumn{1}{c|}{} &
  \multicolumn{1}{l}{\textbf{\begin{tabular}[c]{@{}l@{}}Atomic\\ Evaluation\end{tabular}}} &
  \textbf{\begin{tabular}[c]{@{}l@{}}Propagated\\ Information\end{tabular}} &
  \multicolumn{1}{l|}{\textbf{\begin{tabular}[c]{@{}l@{}}Public\\ Information\end{tabular}}} &
  \multicolumn{1}{l}{\textbf{\begin{tabular}[c]{@{}l@{}}Time\\ Driven\end{tabular}}} &
  \multicolumn{1}{l|}{\textbf{\begin{tabular}[c]{@{}l@{}}Event\\ Driven\end{tabular}}} &
  \multicolumn{1}{c}{} &
  \multicolumn{1}{c}{} \\ \toprule
\multirow{5}{*}{\textbf{VANETs}} 
 & \cite{lu2018bars} 2018 & Entity-Centric & Hybrid & \ding{51} & \ding{51} & \ding{55} & \ding{55} & \ding{51}  & \ding{55} & Statistics \\
 & \cite{luo2019blockchain} 2019& Entity-Centric & Implicit & \ding{51} & \ding{55} & \ding{55} & \ding{55} & \ding{51}  & \ding{51} & Rule \& Logic \\
 & \cite{kudva2021scalable} 2021& Entity-Centric & Implicit & \ding{51} & \ding{51} & \ding{55} & \ding{51} & \ding{51}  & \ding{51} & Statistics \\
 & \cite{youssef2021distributed} 2021 & Entity-Centric & Implicit & \ding{51} & \ding{51} & \ding{55} & \ding{51} & \ding{51} & \ding{51} & Rule \& Logic \\
 & \cite{liu2021behavior} 2021 & Entity-Centric & Implicit & \ding{51} & \ding{55} & \ding{55} & \ding{55} & \ding{51} & \ding{51} & ML-driven  \\
\hline

\multirow{5}{*}{\textbf{IoT}} 
 & \cite{cao2016trust} 2016 & Entity-Centric & Implicit & \ding{51} & \ding{55} & \ding{55} & \ding{55} & \ding{51} & \ding{55} & Rule \& Logic \\
 & \cite{jayasinghe2018machine} 2018 & Entity-centric & Implicit & \ding{51} & \ding{51} & \ding{55} & \ding{55} & \ding{51} & \ding{51} & ML-driven \\
 & \cite{sagar2020trust} 2020 & Entity-centric & Implicit & \ding{51} & \ding{51} & \ding{55} & \ding{55} & \ding{51} & \ding{51} & ML-driven \\
 & \cite{wei2020enhancing} 2020 & Entity-centric & Hybrid & \ding{51} & \ding{51} & \ding{55} & \ding{51} & \ding{51} & \ding{51} & Statistics \\
 & \cite{liu2022semi} 2022 & Entity-centric & Hybrid & \ding{51} & \ding{51} & \ding{55} & \ding{51} & \ding{51} & \ding{51} & Statistics \\
\hline

\multirow{5}{*}{\textbf{MASs}}
 & \cite{zikratov2016dynamic} 2016 & Entity-centric & Implicit & \ding{51} & \ding{51} & \ding{55} & \ding{55} & \ding{51} & \ding{51} & Statistics \\
 & \cite{muller2019trust} 2019 & Entity-centric & Implicit & \ding{51} & \ding{51} & \ding{55} & \ding{51} & \ding{51} & \ding{51} & Rule \& Logic \\
 & \cite{samuel2020trust} 2020 & Entity-centric & Explicit & \ding{51} & \ding{51} & \ding{55} & \ding{51} & \ding{51} & \ding{51} & Statistics \\
 & \cite{cheng2021general} 2021 & Entity-centric & Implicit & \ding{51} & \ding{51} & \ding{55} & \ding{55} & \ding{51} & \ding{51} & Rule \& Logic \\
 & \cite{Wu_2024} 2024 & Entity-centric & Implicit & \ding{51} & \ding{55} & \ding{55} & \ding{55} & \ding{51} & \ding{51} & Statistics \\
\hline

\multirow{4}{*}{\textbf{P2P}}
& \cite{tahta2015gentrust} 2015 & Entity-centric & Hybrid & \ding{51} & \ding{51} & \ding{55} & \ding{55} & \ding{51} & \ding{51} & ML-driven \\
 & \cite{naghizadeh2016c} 2016 & Entity-centric & Explicit & \ding{51} & \ding{51} & \ding{55} & \ding{55} & \ding{51} & \ding{51} & Statistics \\
 & \cite{alkharji2017authenticpeer} 2017 & Combined & Explicit & \ding{51} & \ding{51} & \ding{51} & \ding{51} & \ding{51} & \ding{51} & Graph-based \\
 & \cite{meng2020truetrust} 2020 & Entity-centric & Explicit & \ding{51} & \ding{51} & \ding{55} & \ding{51} & \ding{51} & \ding{51} & Statistics \\ \toprule
\end{tabular}}
\end{table*}
\begin{table*}[t]
\caption{Security Issues of TRMSs in Distributed Autonomous Systems}
\label{tab:security_DAS}
\centering
\renewcommand{\arraystretch}{1.3} 
\begin{tabular}{l|lllllllll}
\bottomrule
\textbf{Domain} &
  \textbf{Reference} &
  \textbf{Sybil Attack} &
  \textbf{\begin{tabular}[c]{@{}l@{}}Pricacy\\ Leakage\end{tabular}} &
  \textbf{\begin{tabular}[c]{@{}l@{}}Cold\\ Start\end{tabular}} &
  \textbf{Whitewashing} &
  \textbf{\begin{tabular}[c]{@{}l@{}}Ballot\\ Stuffing\end{tabular}} &
  \textbf{\begin{tabular}[c]{@{}l@{}}Bad\\ Mouthing\end{tabular}} &
  \textbf{\begin{tabular}[c]{@{}l@{}}On-Off\\ Attack\end{tabular}} &
  \textbf{Collusion} \\ \toprule
\multirow{5}{*}{\textbf{VANETs}} 
 & \cite{lu2018bars} 2018 & \ding{108} & \ding{108} & \ding{109} & \ding{108} & \ding{109} & \ding{108} & \ding{109} & \ding{109} \\
 & \cite{luo2019blockchain} 2019 & \ding{109} & \ding{108} & \ding{109} & \ding{108} & \ding{109} & \ding{108} & \ding{108} & \ding{109} \\
 & \cite{kudva2021scalable} 2021& \ding{108} & \ding{109} & \ding{109} & \ding{108} & \ding{109} & \ding{108} & \ding{109} & \ding{109} \\
 & \cite{youssef2021distributed} 2021 & \ding{109} & \ding{108} & \ding{109} & \ding{109} & \ding{109} & \ding{108} & \ding{109} & \ding{109} \\
 & \cite{liu2021behavior} 2021 & \ding{109} & \ding{108} & \ding{109} & \ding{109} & \ding{109} & \ding{109} & \ding{109} & \ding{109} \\
\hline

\multirow{5}{*}{\textbf{IoT}} 
 & \cite{cao2016trust} 2016 & \ding{109} & \ding{109} & \ding{109} & \ding{109} & \ding{109} & \ding{109} & \ding{109} & \ding{109} \\
 & \cite{jayasinghe2018machine} 2018 & \ding{109} & \ding{109} & \ding{109} & \ding{109} & \ding{109} & \ding{109} & \ding{109} & \ding{109} \\
 & \cite{sagar2020trust} 2020  & \ding{109} & \ding{109} & \ding{109} & \ding{109} & \ding{108} & \ding{108} & \ding{109} & \ding{109} \\
 & \cite{wei2020enhancing} 2020& \ding{109} & \ding{108} & \ding{109} & \ding{108} & \ding{108} & \ding{108} & \ding{109} & \ding{109} \\
 & \cite{liu2022semi} 2022 & \ding{109} & \ding{109} & \ding{109} & \ding{109} & \ding{109} & \ding{109} & \ding{109} & \ding{109} \\
\hline

\multirow{5}{*}{\textbf{MASs}}
 & \cite{zikratov2016dynamic} 2016 & \ding{109} & \ding{109} & \ding{109} & \ding{109} & \ding{109} & \ding{109} & \ding{109} & \ding{109} \\
 & \cite{muller2019trust} 2019 & \ding{108} & \ding{108} & \ding{109} & \ding{109} & \ding{109} & \ding{109} & \ding{109} & \ding{108} \\
 & \cite{samuel2020trust} 2020 & \ding{109} & \ding{108} & \ding{109} & \ding{109} & \ding{109} & \ding{108} & \ding{108} & \ding{109} \\
 & \cite{cheng2021general} 2021 & \ding{109} & \ding{109} & \ding{109} & \ding{109} & \ding{109} & \ding{109} & \ding{109} & \ding{108} \\
 & \cite{Wu_2024} 2024 & \ding{108} & \ding{108} & \ding{109} & \ding{108} & \ding{108} & \ding{108} & \ding{108} & \ding{109} \\
\hline

\multirow{4}{*}{\textbf{P2P}}
 & \cite{tahta2015gentrust} 2015 & \ding{108} & \ding{109} & \ding{109} & \ding{108} & \ding{109} & \ding{109} & \ding{109} & \ding{108} \\
 & \cite{naghizadeh2016c} 2016 & \ding{109} & \ding{109} & \ding{109} & \ding{109} & \ding{109} & \ding{109} & \ding{109} & \ding{109} \\
 & \cite{alkharji2017authenticpeer} 2017 & \ding{109} & \ding{109} & \ding{109} & \ding{109} & \ding{108} & \ding{108} & \ding{108} & \ding{108} \\
 & \cite{meng2020truetrust} 2020 & \ding{108} & \ding{109} & \ding{109} & \ding{108} & \ding{109} & \ding{108} & \ding{108} & \ding{108} \\ \toprule
\end{tabular}
\end{table*}

\section{Evaluation of various TRMSs in Data Sharing}
\label{sec:comparison}
This section presents a systematic analysis of existing TRMSs, applying the proposed taxonomies to synthesize the state-of-the-art across key distributed autonomous systems and digital service ecosystems. The analysis identifies prevailing architectural trends and highlights domain-specific challenges, all viewed through the lens of data sharing. Crucially, this review serves to empirically validate the central gap identified throughout this survey: a predominant focus of existing TRMSs on entity-centric evaluation at the expense of data-centric quality and data consumer compliance.

\subsection{Distributed autonomous systems}
Distributed autonomous systems, such as VANETs, IoT, MAS, and P2P networks, enable independent task execution and cooperative efficiency. However, ensuring entity trustworthiness remains a challenge. TRMSs address this by providing frameworks for trust evaluation, enhancing security and reliability in decentralized environments.

\noindent
\textbf{VANETs:}
In VANETs, vehicles act as both data providers and consumers, sharing time-sensitive and safety-critical information such as traffic conditions and hazard warnings. The primary trust challenge in VANETs is to verify the authenticity of messages and the reliability of participating nodes in a highly dynamic and resource-constrained environment. To address this, a trend in the literature is the adoption of blockchain and other DLTs to provide an immutable and transparent foundation for trust management.

Early efforts, such as the work by Lu \textit{et al.}~\cite{lu2018bars}, exemplify a system with a \textit{Centralized} architecture where a TTP in the form of a Law Enforcement Authority (LEA) performs the final trust calculations. Although leveraging blockchain for storage, the system design is \textit{Domain-Specific} and evaluates entities in a \textit{Homogeneous} manner. As shown in Tab.~\ref{tab:design_DAS}, its evaluation is \textit{Unidirectional}, focusing on the trustworthiness of the data provider while offering robust \textit{Identity} and \textit{Data Protection} through pseudonymity. Subsequent research moved towards \textit{Decentralized} architectures, but often with practical limitations. For example, the models proposed by Luo \textit{et al.}~\cite{luo2019blockchain} and Kudva \textit{et al.}~\cite{kudva2021scalable} rely on Roadside Units (RSUs) for trust computation, creating a functionally centralized process despite the decentralized data storage. The work by Luo \textit{et al.} represents an evolution by introducing a \textit{Bidirectional} and \textit{Role-Specific} evaluation, a better fit for the reciprocal nature of data sharing. More recent approaches show increasing sophistication; Inedjaren \textit{et al.}~\cite{youssef2021distributed} use a \textit{Rule \& Logic} based model (fuzzy logic) for vehicle-level detection of malicious peers, while Liu \textit{et al.}~\cite{liu2021behavior} employ an \textit{ML-driven} model, indicating a shift towards more advanced and truly decentralized computation.

\textit{Discussion:}
Our analysis of TRMSs in VANETs, synthesized from Tab.~\ref{tab:metric_DAS}-\ref{tab:security_DAS}, reveals several key trends and gaps. While there is a clear progression towards \textit{Decentralized} architectures, the trust computation often remains centralized in practice. All reviewed systems are overwhelmingly \textit{Entity-Centric}, focusing on node behavior. Tab.~\ref{tab:metric_DAS} shows that \textit{Data-Centric} metrics like \textit{Authenticity} and \textit{Reliability} are only partially supported, and dimensions such as \textit{Completeness} and \textit{Consistency} are almost entirely overlooked. This validates the central thesis of this survey: the quality of the shared data itself is not treated as a first-class evaluation metric.

Furthermore, given the sensitivity of vehicular data (\textit{e.g.}, location trajectories), the lack of support for \textit{Entity-Centric} metrics like \textit{Compliance} and \textit{Consent} is a major shortcoming. This is compounded by the fact that most models are \textit{Unidirectional}, failing to hold the data consumer accountable. Future research can therefore focus on developing truly decentralized, \textit{Bidirectional} TRMSs that fuse both entity behavior with a richer set of \textit{Data-Centric} metrics. Integrating a quantifiable assessment of consumer \textit{Compliance} is essential for creating a holistic and trustworthy data sharing ecosystem for VANETs.

\noindent
\textbf{IoT networks:}
The IoT represents a massive data sharing ecosystem where heterogeneous devices, acting as data providers, generate vast volumes of data. The primary trust challenges in this domain are twofold: assessing the \textit{QoD}, which can be highly variable due to device heterogeneity and environmental factors~\cite{li2022spatial}, and managing trust computation in a resource-constrained environment.

The analysis of existing literature reveals a strong tendency towards centralized control to manage these challenges. As shown in Tab.~\ref{tab:design_DAS}, the reviewed TRMSs predominantly employ \textit{Centralized} or \textit{Federated} architectures. Early models, like the semantic-based framework by Cao \textit{et al.}~\cite{cao2016trust} and the \textit{ML-driven} model by Jayasinghe \textit{et al.}~\cite{jayasinghe2018machine} and Sagar \textit{et al.}~\cite{sagar2020trust}, offload complex computations to a central authority. These systems are consistently \textit{Entity-Centric}, focusing on service reliability or predicting misbehavior rather than the intrinsic quality of the shared IoT data. Even as more recent models incorporate blockchain for immutable storage, the computational model often remains centralized. The works by Wei \textit{et al.}~\cite{wei2020enhancing} and Liu \textit{et al.}~\cite{liu2022semi}, for instance, use blockchain for decentralized record-keeping but still rely on a central point for trust score aggregation and updates. This trend highlights a pragmatic trade-off in IoT: sacrificing full decentralization to accommodate the limited processing power of end devices.

\textit{Discussion:}
Our analysis of TRMSs in the IoT domain, synthesized from Tab.~\ref{tab:metric_DAS}-\ref{tab:security_DAS}, confirms a consistent design pattern and reveals research gaps. The systems are almost universally \textit{Entity-Centric} and \textit{Unidirectional}, prioritizing the service experience and behavior of the device over the quality of the data it provides. Tab.~\ref{tab:metric_DAS} demonstrates this starkly: none of the reviewed IoT models provide comprehensive support for critical \textit{Data-Centric} metrics such as \textit{Authenticity}, \textit{Integrity}, or \textit{Completeness}. This is a fundamental shortcoming in a domain where the primary value lies in the data itself.

Furthermore, these systems often lack a ``security by design'' approach. As detailed in Tab.~\ref{tab:security_DAS}, none of the surveyed IoT TRMSs explicitly address common threats like \textit{Sybil Attacks} or \textit{Collusion}, and only a few consider \textit{Whitewashing}. Future research can therefore pivot towards developing TRMSs for IoT that are fundamentally \textit{data-centric}, using the rich metrics from Sec.~\ref{subsec:data_centric metrics} to directly assess sensor data quality. There is a need for lightweight, secure-by-design, and truly decentralized evaluation frameworks that can operate effectively at the massive scale of the IoT data sharing ecosystem.

\noindent
\textbf{MASs:}
MASs are inherently data-sharing ecosystems, where numerous intelligent agents collaboratively exchange information to achieve a shared objective~\cite{wu2024proof}. Trust is paramount, as the collective goal can be compromised by a single malicious or faulty agent providing false data. The literature on TRMSs for MASs, therefore, focuses heavily on verifying agent behavior to ensure reliable cooperation.

Our analysis of the literature, synthesized in Tab.~\ref{tab:design_DAS}, reveals a dominant design pattern: TRMSs for MASs are typically \textit{Decentralized}, \textit{Domain-Specific}, and evaluate agents in a \textit{Homogeneous} manner. This approach, seen in the works of Zikratov \textit{et al.}~\cite{zikratov2016dynamic}, Müller \textit{et al.}~\cite{muller2019trust}, and Samuel \textit{et al.}~\cite{samuel2020trust}, supports the autonomous nature of agents. These systems are almost universally \textit{Entity-Centric}, employing \textit{Rule \& Logic} based models like subjective logic to detect and isolate misbehaving agents. An interesting outlier is the framework by Cheng \textit{et al.}~\cite{cheng2021general}, which uses a \textit{Centralized} yet \textit{Domain-General} architecture, demonstrating its applicability across multiple MAS scenarios. As agents often operate in the physical world, recent models have introduced more advanced verification methods. The framework by Wu \textit{et al.}~\cite{Wu_2024}, for example, uses proof-of-location as a novel \textit{Implicit Signal} to objectively verify task completion, moving beyond purely reputation-based evidence.

\textit{Discussion:}
The analysis of TRMSs in MASs, summarized in Tab.~\ref{tab:metric_DAS}-\ref{tab:security_DAS}, validates this survey's central thesis. The evaluation is overwhelmingly \textit{Entity-Centric}, designed to detect misbehavior. As shown in Tab.~\ref{tab:metric_DAS}, while some systems support basic \textit{Data-Centric} metrics like \textit{Authenticity} and \textit{Integrity} as a byproduct of behavioral verification, they lack a comprehensive framework to assess the full spectrum of data quality. This is a critical gap, as trustworthy behavior is meaningless if the data being shared is inaccurate or incomplete.

Furthermore, the models are almost all \textit{Unidirectional}, assessing agents as information providers but not as consumers. In a truly collaborative environment, a \textit{Bidirectional} model is needed for mutual accountability. Resource constraints also appear to limit the adoption of DLTs, with only a few reviewed systems~\cite{samuel2020trust,Wu_2024} using it for immutable storage to defend against threats like \textit{Whitewashing}, as noted in Tab.~\ref{tab:evaluation_DAS}. Future research can therefore focus on developing lightweight, \textit{Bidirectional} TRMSs for MASs that explicitly integrate a rich set of \textit{Data-Centric} metrics to ensure that the information shared between agents is as trustworthy as the agents themselves.

\noindent
\textbf{P2P network:}
P2P networks represent one of the earliest forms of large-scale, decentralized data sharing. The primary trust challenges in this domain are twofold: verifying the quality and authenticity of files shared by unknown peers and mitigating anti-social behaviors such as freeriding. The literature reflects these challenges, showing a strong focus on decentralized governance and behavioral incentives.

As synthesized in Tab.~\ref{tab:design_DAS}, TRMSs in this domain are predominantly built on a \textit{Decentralized} architecture to reflect the native structure of P2P systems. Early and common models, such as those by Tahta \textit{et al.}~\cite{tahta2015gentrust} and Naghizadeh \textit{et al.}~\cite{naghizadeh2016c}, are \textit{Entity-Centric}, applying \textit{ML-driven} (genetic programming) or \textit{Statistical} models to evaluate peer behavior and ensure fairness. A notable evolution is seen in the work of Alkharji \textit{et al.}~\cite{alkharji2017authenticpeer}, which introduces a \textit{Combined} evaluation model that assesses both the entity and the data. As shown in Tab.~\ref{tab:evaluation_DAS}, it is the only reviewed P2P system to employ a \textit{Graph-based} computational model, leveraging the network topology to compute trust. While most systems are decentralized, the approach in~\cite{meng2020truetrust} represents a \textit{Centralized} model designed to enhance robustness by emphasizing the requester's own feedback to filter malicious inputs.

\textit{Discussion:}
Our analysis of TRMSs in P2P networks reveals a domain with a stronger, albeit still underdeveloped, focus on the shared data compared to other distributed systems. As shown in Tab.~\ref{tab:metric_DAS}, systems like~\cite{meng2020truetrust,alkharji2017authenticpeer} support \textit{Data-Centric} metrics like \textit{Authenticity}, \textit{Integrity}, and \textit{Validity}. However, the evaluation methods are not sufficiently robust, failing to account for the full spectrum of QoD metrics needed for high-stakes data sharing.

The primary gaps lie in privacy and accountability for data handling. Tab.~\ref{tab:design_DAS} shows a complete lack of advanced \textit{Privacy-Preserving Techniques}; none of the reviewed systems provide \textit{Data Protection} or \textit{Computational Privacy}. Furthermore, Tab.~\ref{tab:metric_DAS} confirms that no systems consider \textit{Compliance} or \textit{Consent}, a critical oversight as P2P architectures are increasingly explored for more sensitive data sharing applications. Future work can build on the existing data-centric foundation of P2P TRMSs by incorporating a richer set of QoD metrics and, crucially, integrating robust privacy-preserving technologies and compliance monitoring frameworks to ensure these systems are viable for the next generation of sensitive, decentralized data exchange.

\begin{table*}
\centering
\caption{Comparison of Trust Evaluation Metrics in Digital Service Ecosystems}
\label{tab:metric_DSE}
\begin{threeparttable}
\begingroup

\renewcommand{\arraystretch}{1.5} 
\begin{tabular}{l|lllllllllllllllllll}
\bottomrule
\multirow{2}{*}{\textbf{Domain}} &
\multirow{2}{*}{\textbf{Reference}} &
\multicolumn{12}{|c}{\textbf{Data}\tnote{I}} &  
\multicolumn{6}{|c}{\textbf{Entity}\tnote{II}}  
\\
\cline{3-14} \cline{15-20}  
 &  & \multicolumn{1}{|l}{\textbf{Au}} & \textbf{Ra} & \textbf{Va} & \textbf{In} & \textbf{Ac} & \textbf{Cp} & \textbf{Cs} & \textbf{Ti} & \textbf{Rl} & \textbf{Pr} & \textbf{Tr} & \textbf{Un} & \multicolumn{1}{|l}{\textbf{Cl}} & \textbf{Tp} & \textbf{Ao} & \textbf{Se} & \textbf{Co} & \textbf{Re} \\ 
\toprule
  
\multirow{3}{*}{\textbf{Healthcare}} 
                        & \cite{meng2018towards} 2018 & \ding{108} & \ding{109} & \ding{108} & \ding{109} & \ding{109} & \ding{109} & \ding{109} & \ding{109} & \ding{109} & \ding{109} & \ding{108} & \ding{109} & \ding{109} & \ding{109} & \ding{108} & \ding{109} & \ding{109} & \ding{108} \\ 
                        & \cite{khalfaoui2020stochastic} 2020 & \ding{108} & \ding{108} & \ding{109} & \ding{108} & \ding{109} & \ding{109} & \ding{109} & \ding{109} & \ding{109} & \ding{109} & \ding{109} & \ding{109} & \ding{109} & \ding{109} & \ding{108} & \ding{108} & \ding{109} & \ding{108} \\ 
                        & \cite{bhan2023blockchain} 2023 & \ding{108} & \ding{108} & \ding{108} & \ding{108} & \ding{108} & \ding{109} & \ding{109} & \ding{108} & \ding{109} & \ding{109} & \ding{108} & \ding{109} & \ding{109} & \ding{109} & \ding{108} & \ding{108} & \ding{109} & \ding{108} \\  \hline
\multirow{4}{*}{\textbf{Fog/Edge}} 
                        & \cite{kang2020reliable} 2020 & \ding{109} & \ding{109} & \ding{108} & \ding{109} & \ding{109} & \ding{109} & \ding{109} & \ding{108} & \ding{109} & \ding{109} & \ding{109} & \ding{109} & \ding{109} & \ding{109} & \ding{109} & \ding{109} & \ding{109} & \ding{108} \\ 
                        & \cite{wang2020reliable} 2020 & \ding{109} & \ding{109} & \ding{109} & \ding{109} & \ding{109} & \ding{109} & \ding{109} & \ding{109} & \ding{108} & \ding{109} & \ding{109} & \ding{109} & \ding{109} & \ding{109} & \ding{109} & \ding{109} & \ding{109} & \ding{108} \\ 
                        & \cite{ogundoyin2021trust} 2021 & \ding{109} & \ding{109} & \ding{109} & \ding{108} & \ding{109} & \ding{109} & \ding{109} & \ding{109} & \ding{109} & \ding{109} & \ding{109} & \ding{109} & \ding{109} & \ding{109} & \ding{109} & \ding{108} & \ding{109} & \ding{108} \\ 
                        & \cite{latif2022novel} 2022 & \ding{109} & \ding{109} & \ding{109} & \ding{109} & \ding{109} & \ding{109} & \ding{109} & \ding{109} & \ding{109} & \ding{109} & \ding{109} & \ding{109} & \ding{109} & \ding{109} & \ding{109} & \ding{109} & \ding{109} & \ding{108} \\ 
                        \hline 
\multirow{2}{*}{\textbf{Crowdsource}} 
                        & \cite{yu2020crowdr} 2020 & \ding{109} & \ding{109} & \ding{109} & \ding{109} & \ding{108} & \ding{109} & \ding{109} & \ding{108} & \ding{109} & \ding{109} & \ding{109} & \ding{109} & \ding{109} & \ding{109} & \ding{109} & \ding{109} & \ding{109} & \ding{108} \\ 
                        & \cite{wang2020blockchain} 2020 & \ding{109} & \ding{109} & \ding{109} & \ding{108} & \ding{109} & \ding{109} & \ding{109} & \ding{109} & \ding{109} & \ding{109} & \ding{109} & \ding{109} & \ding{109} & \ding{109} & \ding{109} & \ding{108} & \ding{109} & \ding{108} \\ 
                        \hline
\multirow{2}{*}{\textbf{Social Nets}} 
                        & \cite{lin2020guardian} 2020 & \ding{109} & \ding{109} & \ding{109} & \ding{109} & \ding{109} & \ding{109} & \ding{109} & \ding{109} & \ding{109} & \ding{109} & \ding{109} & \ding{109} & \ding{109} & \ding{109} & \ding{109} & \ding{109} & \ding{109} & \ding{108} \\ 
                        & \cite{aghdam2020uncertainty} 2020 & \ding{109} & \ding{109} & \ding{109} & \ding{109} & \ding{109} & \ding{109} & \ding{109} & \ding{109} & \ding{109} & \ding{109} & \ding{109} & \ding{109} & \ding{109} & \ding{109} & \ding{109} & \ding{109} & \ding{109} & \ding{108} \\ 
                        \hline 
      
\multirow{2}{*}{\textbf{Data Market}}
& \cite{chowdhury2019trust} 2019 & \ding{109} & \ding{109} & \ding{109} & \ding{108} & \ding{108} & \ding{109} & \ding{109} & \ding{109} & \ding{109} & \ding{109} & \ding{108} & \ding{109} & \ding{109} & \ding{108} & \ding{108} & \ding{108} & \ding{108} & \ding{108} \\ 
 & \cite{hesse2020takeaway} 2020 & \ding{108} & \ding{108} & \ding{109} & \ding{108} & \ding{109} & \ding{109} & \ding{108} & \ding{109} & \ding{108} & \ding{109} & \ding{109} & \ding{109} & \ding{109} & \ding{108} & \ding{109} & \ding{109} & \ding{109} & \ding{108} \\ 
   
\toprule
\end{tabular}%
\endgroup
 \begin{tablenotes}
 \item  \ding{109}: not supported;
        \ding{108}: supported.
 \item I: \textbf{Au:} Authenticity, \textbf{Ra:} Reliability, \textbf{Va:} Validity,  \textbf{In:} Integrity, \textbf{Ac:} Accuracy, \textbf{Cp:} Completeness, \textbf{Cs:} Consistency, \textbf{Ti:} Timeliness \textbf{Rl:} Relevance, \textbf{Pr:} Precision, \textbf{Tr:} Traceability, \textbf{Un:} Uniqueness.
  \item II: \textbf{Cl:} Compliance, \textbf{Tp:} Transparency, \textbf{Ao:} Accountability, \textbf{Se:} Security, \textbf{Co:} Consent, \textbf{Re:} Reputation.
 \end{tablenotes}
 \end{threeparttable}
\end{table*}

\begin{table*}[t]
\caption{System Design of TRMSs in Digital Service Ecosystems}
\label{tab:design_DSE}
\centering
\resizebox{\textwidth}{!}{%
\renewcommand{\arraystretch}{1.5} 
\begin{tabular}{l|lllp{0.07\linewidth}llccc}
\bottomrule
\multirow{3}{*}{\textbf{Domain}} &
  \multicolumn{1}{l}{\multirow{3}{*}{\textbf{Reference}}} &
  \multirow{3}{*}{\textbf{Architecture}} &
  \multirow{3}{*}{\textbf{\begin{tabular}[c]{@{}l@{}}Evaluation\\ Granularity\end{tabular}}} &
  \multirow{3}{*}{\textbf{\begin{tabular}[c]{@{}l@{}}Domain\\ Specificity\end{tabular}}} &
  \multirow{3}{*}{\textbf{\begin{tabular}[c]{@{}l@{}}Evaluation\\ Adaptability\end{tabular}}} &
  \multicolumn{1}{l|}{\multirow{3}{*}{\textbf{\begin{tabular}[c]{@{}l@{}}Evaluation\\ Directionality\end{tabular}}}} &
  \multicolumn{3}{c}{\textbf{Privacy-Preserving Techniques}} \\ \cline{8-10} 
 &
  \multicolumn{1}{l}{} &
   &
   &
   &
   &
  \multicolumn{1}{l|}{} &
  \textbf{\begin{tabular}[c]{@{}l@{}}Identity\\ Protection\end{tabular}} &
  \textbf{\begin{tabular}[c]{@{}l@{}}Data\\ Protection\end{tabular}} &
  \textbf{\begin{tabular}[c]{@{}l@{}}Computational\\ Privacy\end{tabular}} \\ \toprule
\multirow{3}{*}{\textbf{Healthcare}} 
 & \cite{meng2018towards} 2018
   & Centralized  
   & Homogeneous
   & Specific
   & Dynamic
   & Unidirectional
   & \ding{55}
   & \ding{55}
   & \ding{55} \\
 & \cite{khalfaoui2020stochastic} 2020
   & Decentralized
   & Homogeneous
   & Specific
   & Dynamic
   & Unidirectional
   & \ding{55}
   & \ding{51}
   & \ding{55} \\
 & \cite{bhan2023blockchain} 2023
   & Centralized
   & Homogeneous
   & Specific
   & Dynamic
   & Unidirectional
   & \ding{55}
   & \ding{51}   
   & \ding{55} \\ 
\hline

\multirow{4}{*}{\textbf{Fog/Edge}} 
 & \cite{kang2020reliable} 2020
   & Decentralized
   & Role-Specific
   & Specific
   & Dynamic
   & Unidirectional
   & \ding{51}
   & \ding{51}
   & \ding{55} \\
 & \cite{wang2020reliable} 2020
   & Centralized
   & Role-Specific
   & Specific
   & Dynamic
   & Unidirectional
   & \ding{51}
   & \ding{55}
   & \ding{55} \\
 & \cite{ogundoyin2021trust} 2021
   & Decentralized
   & Role-Specific
   & Specific
   & Dynamic
   & Bidirectional
   & \ding{55}
   & \ding{55}
   & \ding{55} \\
 & \cite{latif2022novel} 2022
   & Centralized
   & Homogeneous
   & Specific
   & Dynamic
   & Unidirectional
   & \ding{55}
   & \ding{55}
   & \ding{55} \\
\hline

\multirow{2}{*}{\textbf{Crowdsource}}
 & \cite{yu2020crowdr} 2020
   & Decentralized
   & Homogeneous
   & Specific
   & Dynamic
   & Unidirectional
   & \ding{51}
   & \ding{51}
   & \ding{55} \\
 & \cite{wang2020blockchain} 2020
   & Decentralized
   & Homogeneous
   & Specific
   & Dynamic
   & Unidirectional
   & \ding{55}
   & \ding{51}
   & \ding{51} \\
\hline

\multirow{2}{*}{\textbf{Social Nets}}
 & \cite{lin2020guardian} 2020
   & Centralized
   & Homogeneous
   & Specific
   & Dynamic
   & Unidirectional
   & \ding{55}
   & \ding{55}
   & \ding{55} \\
 & \cite{aghdam2020uncertainty} 2020
   & Centralized
   & Homogeneous
   & Specific
   & Dynamic
   & Unidirectional
   & \ding{55}
   & \ding{55}
   & \ding{55} \\ \hline
   
\multirow{2}{*}{\textbf{Data Market}}
 & \cite{chowdhury2019trust} 2019
   & Centralized
   & Role-Specific
   & Specific
   & Dynamic
   & Bidirectional
   & \ding{51}
   & \ding{51}
   & \ding{55} \\
 & \cite{hesse2020takeaway} 2020
   & Centralized
   & Homogeneous
   & Specific
   & Dynamic
   & Unidirectional
   & \ding{55}
   & \ding{55}
   & \ding{55} \\\toprule
\end{tabular}
}
\end{table*}
\begin{table*}[]
\caption{Trust Evaluation Framework of TRMSs in Digital Service Ecosystems}
\label{tab:evaluation_DSE}
\centering
\resizebox{\textwidth}{!}{%
\renewcommand{\arraystretch}{1.5} 
\begin{tabular}{l|llllllllll}
\bottomrule
\multicolumn{1}{c|}{\multirow{3}{*}{\textbf{Domain}}} &
  \multicolumn{1}{c}{\multirow{3}{*}{\textbf{Reference}}} &
  \multicolumn{1}{c}{\multirow{3}{*}{\textbf{\begin{tabular}[c]{@{}c@{}}Evaluation\\ Metrics\end{tabular}}}} &
  \multicolumn{1}{c|}{\multirow{3}{*}{\textbf{\begin{tabular}[c]{@{}c@{}}Trust\\ Signals\end{tabular}}}} &
  \multicolumn{3}{c|}{\textbf{Data Sources}} &
  \multicolumn{2}{c|}{\textbf{\begin{tabular}[c]{@{}c@{}}Aggregation\\ Strategies\end{tabular}}} &
  \multicolumn{1}{l}{\multirow{3}{*}{\textbf{\begin{tabular}[l]{@{}l@{}}Pattern\\ Detection\end{tabular}}}} &
  \multicolumn{1}{l}{\multirow{3}{*}{\textbf{\begin{tabular}[l]{@{}l@{}}Computational\\ Models\end{tabular}}}} \\ \cline{5-9}
\multicolumn{1}{c|}{} &
  \multicolumn{1}{c}{} &
  \multicolumn{1}{c}{} &
  \multicolumn{1}{c|}{} &
  \multicolumn{1}{c}{\textbf{\begin{tabular}[c]{@{}l@{}}Atomic\\ Evaluation\end{tabular}}} &
  \textbf{\begin{tabular}[c]{@{}l@{}}Propagated\\ Information\end{tabular}} &
  \multicolumn{1}{l|}{\textbf{\begin{tabular}[c]{@{}l@{}}Public\\ Information\end{tabular}}} &
  \multicolumn{1}{l}{\textbf{\begin{tabular}[c]{@{}l@{}}Time\\ Driven\end{tabular}}} &
  \multicolumn{1}{l|}{\textbf{\begin{tabular}[c]{@{}l@{}}Event\\ Driven\end{tabular}}} &
  \multicolumn{1}{l}{} &
  \multicolumn{1}{l}{} \\ \toprule
\multirow{3}{*}{\textbf{Healthcare}} 
& \cite{meng2018towards} 2018 & Entity-centric & Implicit & \ding{51} & \ding{55} & \ding{55} & \ding{51} & \ding{51} & \ding{51} & Rule \& Logic \\
 & \cite{khalfaoui2020stochastic} 2020 & Entity-centric & Implicit & \ding{51} & \ding{55} & \ding{55} & \ding{51} & \ding{51} & \ding{51} & Rule \& Logic \\
 & \cite{bhan2023blockchain} 2023 & Entity-centric & Hybrid & \ding{51} & \ding{51} & \ding{55} & \ding{51} & \ding{51} & \ding{51} & Statistics \\
\hline

\multirow{4}{*}{\textbf{Fog/Edge}} 
& \cite{kang2020reliable} 2020 & Entity-centric & Implicit & \ding{51} & \ding{55} & \ding{55} & \ding{51} & \ding{51} & \ding{51} & Statistics \\
 & \cite{wang2020reliable} 2020 & Entity-centric & Hybrid & \ding{51} & \ding{51} & \ding{55} & \ding{51} & \ding{51} & \ding{51} & Statistics \\
 & \cite{ogundoyin2021trust} 2021 & Entity-centric & Hybrid & \ding{51} & \ding{51} & \ding{55} & \ding{55} & \ding{51} & \ding{51} & Rule \& Logic \\
 & \cite{latif2022novel} 2022 & Entity-centric & Implicit & \ding{51} & \ding{51} & \ding{55} & \ding{51} & \ding{51} & \ding{51} & Statistics \\
\hline

\multirow{2}{*}{\textbf{Crowdsource}}
& \cite{yu2020crowdr} 2020 & Entity-centric & Explicit & \ding{51} & \ding{51} & \ding{55} & \ding{51} & \ding{51} & \ding{51} & Statistics \\
 & \cite{wang2020blockchain} 2020 & Entity-centric & Hybrid & \ding{51} & \ding{51} & \ding{55} & \ding{51} & \ding{51} & \ding{51} & Rule \& Logic \\
\hline

\multirow{2}{*}{\textbf{Social Nets}}
& \cite{lin2020guardian} 2020 & Entity-centric & Implicit & \ding{55} & \ding{51} & \ding{55} & \ding{55} & \ding{55} & \ding{55} & ML-driven \\
 & \cite{aghdam2020uncertainty} 2020 & Entity-centric & Explicit & \ding{51} & \ding{51} & \ding{55} & \ding{55} & \ding{51} & \ding{55} & Rule \& Logic \\ \hline
      
\multirow{2}{*}{\textbf{Data Market}}
& \cite{chowdhury2019trust} 2019 & Combined & Hybrid & \ding{51} & \ding{51} & \ding{55} & \ding{51} & \ding{51} & \ding{51} & Statistics \\
 & \cite{hesse2020takeaway} 2020 & Entity-centric & Explicit & \ding{51} & \ding{51} & \ding{55} & \ding{51} & \ding{51} & \ding{55}& Statistics \\
   
   \toprule
\end{tabular}}
\end{table*}
\begin{table*}[t]
\caption{Security Issues of TRMSs in Digital Service Ecosystems}
\label{tab:security_DSE}
\centering
\renewcommand{\arraystretch}{1.3} 
\begin{tabular}{l|lllllllll}
\bottomrule
\textbf{Domain} &
  \textbf{Reference} &
  \textbf{Sybil Attack} &
  \textbf{\begin{tabular}[c]{@{}l@{}}Pricacy\\ Leakage\end{tabular}} &
  \textbf{\begin{tabular}[c]{@{}l@{}}Cold\\ Start\end{tabular}} &
  \textbf{Whitewashing} &
  \textbf{\begin{tabular}[c]{@{}l@{}}Ballot\\ Stuffing\end{tabular}} &
  \textbf{\begin{tabular}[c]{@{}l@{}}Bad\\ Mouthing\end{tabular}} &
  \textbf{\begin{tabular}[c]{@{}l@{}}On-Off\\ Attack\end{tabular}} &
  \textbf{Collusion} \\ \toprule
\multirow{3}{*}{\textbf{Healthcare}} 
 & \cite{meng2018towards} 2018 &
\ding{109} & \ding{108} & \ding{109} & \ding{109} & \ding{109} & \ding{109} & \ding{109} & \ding{109} \\
 & \cite{khalfaoui2020stochastic} 2020 &
\ding{109} & \ding{109} & \ding{109} & \ding{109} & \ding{109} & \ding{109} & \ding{108} & \ding{109} \\
 & \cite{bhan2023blockchain} 2023 &
\ding{109} & \ding{108} & \ding{108} & \ding{109} & \ding{108} & \ding{108} & \ding{108} & \ding{108} \\ 
\hline

\multirow{4}{*}{\textbf{Fog/Edge}} 
 & \cite{kang2020reliable} 2020 &
\ding{109} & \ding{109} & \ding{109} & \ding{109} & \ding{109} & \ding{109} & \ding{109} & \ding{108} \\
 & \cite{wang2020reliable} 2020 &
\ding{109} & \ding{109} & \ding{109} & \ding{109} & \ding{108} & \ding{108} & \ding{108} & \ding{108} \\
 & \cite{ogundoyin2021trust} 2021 &
\ding{109} & \ding{109} & \ding{109} & \ding{109} & \ding{108} & \ding{108} & \ding{108} & \ding{109} \\
 & \cite{latif2022novel} 2022 &
\ding{109} & \ding{109} & \ding{109} & \ding{109} & \ding{109} & \ding{109} & \ding{109} & \ding{109} \\
\hline

\multirow{2}{*}{\textbf{Crowdsource}}
 & \cite{yu2020crowdr} 2020 &
\ding{109} & \ding{109} & \ding{109} & \ding{109} & \ding{109} & \ding{109} & \ding{109} & \ding{108} \\
 & \cite{wang2020blockchain} 2020 &
\ding{109} & \ding{108} & \ding{109} & \ding{109} & \ding{109} & \ding{109} & \ding{109} & \ding{109} \\
\hline

\multirow{2}{*}{\textbf{Social Nets}}
 & \cite{lin2020guardian} 2020 &
\ding{109} & \ding{109} & \ding{109} & \ding{109} & \ding{109} & \ding{109} & \ding{109} & \ding{109} \\
 & \cite{aghdam2020uncertainty} 2020 &
\ding{109} & \ding{109} & \ding{109} & \ding{109} & \ding{109} & \ding{109} & \ding{109} & \ding{109} \\ \hline

\multirow{2}{*}{\textbf{Data Market}}
& \cite{chowdhury2019trust} 2019 & \ding{109} & \ding{108} & \ding{109} & \ding{109} & \ding{109} & \ding{109} & \ding{109} & \ding{109} \\ 
 & \cite{hesse2020takeaway} 2020 & \ding{109} & \ding{109} & \ding{108} & \ding{108} & \ding{108} & \ding{108} & \ding{109} & \ding{109} \\

\toprule
\end{tabular}
\end{table*}

\subsection{Digital service ecosystems}
TRMSs are also widely adopted in various digital service ecosystems to enhance user experiences, as trust values can help identify the most reliable entities for interaction. In this subsection, we evaluate several TRMSs in the domains of healthcare, fog/edge computing, crowdsourcing, and social networks. The comparison of these TRMSs is illustrated in Tab.~\ref{tab:metric_DSE} and \ref{tab:security_DSE}.

\noindent
\textbf{Healthcare:}
The healthcare domain represents a high-stakes data sharing environment where the trustworthiness of both the data and the participating entities is paramount for patient safety and regulatory compliance. The primary trust challenge in this context is preventing malicious insider attacks and ensuring the integrity of sensitive health data. The literature reflects this focus, with proposed TRMSs designed primarily as internal security mechanisms rather than frameworks for open data exchange.

Our analysis of the literature, synthesized in Tab.~\ref{tab:design_DSE}, reveals that existing TRMSs for healthcare are mostly built on \textit{Centralized} architectures. The models proposed by Meng \textit{et al.}~\cite{meng2018towards} and the work in~\cite{bhan2023blockchain}, for example, use centralized control to manage trust within a single organizational boundary (\textit{e.g.}, a hospital's software-defined network). These systems are consistently \textit{Entity-Centric} and \textit{Unidirectional}, designed to monitor the behavior of internal nodes and detect threats like insider attacks. While Khalfaoui \textit{et al.}~\cite{khalfaoui2020stochastic} introduces a \textit{Decentralized} model using blockchain to address \textit{On-Off Attacks}, the broader trend points towards centralized governance. The evaluation framework for these systems, as shown in Tab.~\ref{tab:evaluation_DSE}, relies on \textit{Implicit} or \textit{Hybrid} trust signals and employs \textit{Rule \& Logic} or \textit{Statistical} computational models to classify node behavior.

\textit{Discussion:}
The analysis of healthcare TRMSs, summarized in Tab.~\ref{tab:metric_DSE}-\ref{tab:security_DSE}, confirms that the field is nascent and narrowly focused. The consistent use of \textit{Centralized}, \textit{Unidirectional}, and \textit{Entity-Centric} designs is logical for detecting insider threats within a single organization, but it is insufficient for the challenges of large-scale, cross-organization data sharing. Tab.~\ref{tab:metric_DSE} highlights another gap: a near-total lack of support for \textit{Data-Centric} metrics. While some models consider data \textit{Authenticity} and \textit{Reliability}, the full spectrum of QoD is ignored, a significant oversight where data quality can directly impact clinical outcomes.

Furthermore, for data sharing to succeed, the trustworthiness of the data consumer is as important as that of the provider. The \textit{Unidirectional} models fail to hold consumers accountable for their data handling practices. Future research can therefore focus on developing \textit{Bidirectional} TRMSs for healthcare that integrate a comprehensive, \textit{Data-Centric} evaluation of QoD. To facilitate trustworthy cross-organizational data exchange, these future systems can also incorporate robust validation of consumer \textit{Compliance} with data sharing agreements and explore the use of advanced \textit{Computational Privacy} techniques to protect sensitive patient information.

\noindent
\textbf{Fog/edge computing:}
In modern data sharing architectures, fog and edge computing serve as an intermediate layer, processing data closer to its source to reduce latency and enhance efficiency. The primary trust challenge in this domain is ensuring the reliability and quality of services provided by a heterogeneous and dynamic set of edge nodes. Consequently, the TRMS literature in this area is heavily focused on evaluating service-level performance.

Our analysis of existing works, synthesized in Tab.~\ref{tab:design_DSE}, shows a preference for \textit{Decentralized} architectures suited to the distributed nature of edge computing, as seen in the models by Kang \textit{et al.}~\cite{kang2020reliable} and Ogundoyin \textit{et al.}~\cite{ogundoyin2021trust}. A key characteristic of these systems is their \textit{Role-Specific} evaluation granularity; they are designed to differentiate between various actors, such as workers in a federated learning task or service providers and requesters. An advancement is demonstrated by Ogundoyin \textit{et al.}, which is the only reviewed model in this domain to employ a \textit{Bidirectional} evaluation framework, allowing for mutual assessment between providers and consumers. This model is also notable for its resilience against \textit{Bad Mouthing} and \textit{Ballot Stuffing} attacks, as shown in Tab.~\ref{tab:security_DSE}, and its use of a \textit{Rule \& Logic} based computational model (fuzzy logic) to handle the uncertainty inherent in mobile environments.

\textit{Discussion:}
The analysis of TRMSs in fog/edge computing, summarized in Tab.~\ref{tab:metric_DSE}-\ref{tab:security_DSE}, reveals a domain that has matured in terms of service reliability but remains underdeveloped for high-quality data sharing. The evaluation in these systems is universally \textit{Entity-Centric}, designed to answer the question, ``Is the service reliable?'' by measuring operational quality of service (QoS) parameters like latency and throughput, as exemplified by Latif \textit{et al.}~\cite{latif2022novel}. 

This focus creates a gap, which is clearly visible in Tab.~\ref{tab:metric_DSE}: a near-total absence of support for \textit{Data-Centric} metrics. None of the reviewed systems assess the \textit{Authenticity}, \textit{Accuracy}, or \textit{Completeness} of the data being processed at the edge. This narrow focus on QoS limits their applicability in large-scale data sharing ecosystems, where the intrinsic quality of the data is often more important than the performance of the service delivering it. Furthermore, as shown in Tab.~\ref{tab:evaluation_DSE}, while some systems provide basic \textit{Identity Protection}, none employ advanced \textit{Computational Privacy} techniques, an oversight for a layer that processes potentially sensitive data. Future research can therefore engineer a paradigm shift, designing TRMSs for the edge that move beyond QoS to incorporate a rich set of \textit{Data-Centric} metrics. This will ensure that the trustworthiness of the data itself is evaluated with the same rigor as the services that process it.

\noindent
\textbf{Crowdsourcing:}
Crowdsourcing represents a powerful paradigm for large-scale data sharing, typically between individual workers (data providers) and organizations (data consumers). The primary trust challenges in this domain are ensuring the privacy of a large, distributed workforce and guaranteeing the reliability of their contributions against malicious behaviors like reputation manipulation. Consequently, the TRMS literature in this area showcases sophisticated, secure, and privacy-aware architectures.

The state-of-the-art in crowdsourcing TRMSs, exemplified by the work of Yu \textit{et al.}~\cite{yu2020crowdr} and Wang \textit{et al.}~\cite{wang2020blockchain}, is characterized by advanced \textit{Decentralized} architectures. As detailed in Tab.~\ref{tab:evaluation_DSE}, these systems are designed to be \textit{Domain-Specific} and \textit{Unidirectional}, focusing on the trustworthiness of the worker. They leverage blockchain to ensure transparency and mitigate attacks, and as Tab.~\ref{tab:design_DSE} shows, they integrate \textit{Privacy-Preserving Techniques} for both \textit{Identity} and \textit{Data Protection} as a core design principle. The evaluation frameworks, detailed in Tab.~\ref{tab:evaluation_DSE}, are \textit{Entity-Centric}, using \textit{Hybrid} or \textit{Explicit} signals to evaluate worker behavior. These systems often incorporate novel technologies like Trusted Execution Environments (TEEs) and Proof of Trust (PoT) consensus mechanisms to enhance security and scalability~\cite{wang2020blockchain}.

\textit{Discussion:}
While TRMSs in crowdsourcing are architecturally advanced, our analysis reveals that their evaluation frameworks are underdeveloped for high-quality data sharing. Despite their sophistication, the models remain limited by their \textit{Entity-Centric} and \textit{Unidirectional} nature. Tab.~\ref{tab:metric_DSE} illustrates this gap: the systems almost completely ignore \textit{Data-Centric} metrics. They are designed to assess the reliability of the worker, but not the intrinsic \textit{Accuracy}, \textit{Completeness}, or \textit{Relevance} of the data they provide. 

This focus on the data provider at the expense of the data is a shortcoming. Furthermore, the \textit{Unidirectional} model provides no mechanism to assess the behavior of the requester (the data consumer), which is essential for a balanced ecosystem. For crowdsourcing to mature into a truly trustworthy data-sharing framework, future research can shift its focus. It is imperative to develop TRMSs that integrate comprehensive \textit{Data-Centric} metrics to assess the quality of the crowdsourced data itself. Additionally, adopting \textit{Bidirectional} evaluation models is essential to hold data consumers accountable for their actions, thereby creating a more equitable and trustworthy environment for all participants.

\noindent
\textbf{Social Networks:}
Social networks are massive, user-driven data sharing platforms where the shared content ranges from personal updates to news and public information. The primary trust challenge in this domain is to help users assess the reliability of information and the trustworthiness of other users within a vast, interconnected graph, particularly in the presence of misinformation. The TRMS literature for this domain, therefore, focuses on developing computationally efficient and accurate models for trust prediction at scale.

The reviewed systems exemplify the state-of-the-art, which, as shown in Tab.~\ref{tab:design_DSE}, is characterized by \textit{Centralized} architectures that are natural for platform-based social media. These models are \textit{Entity-Centric} and \textit{Unidirectional}, designed to evaluate the trustworthiness of users. The Guardian framework by Lin \textit{et al.}~\cite{lin2020guardian}, for instance, is a \textit{ML-driven} model that leverages Graph Convolutional Networks (GCNs) to efficiently process the network's topology. As detailed in Tab.~\ref{tab:evaluation_DSE}, it is unique in that its evaluation relies entirely on \textit{Propagated Information} from the social graph rather than atomic, event-based signals. In contrast, the model by Aghdam \textit{et al.}~\cite{aghdam2020uncertainty} uses a \textit{Rule \& Logic} based approach (subjective logic) to capture the nuances of social trust and distrust based on \textit{Explicit Signals}.

\textit{Discussion:}
Our analysis reveals that TRMSs for social networks are computationally advanced but conceptually narrow. Their primary focus is on the efficiency and accuracy of predicting user reputation within a closed, \textit{Entity-Centric} paradigm. This creates a gap, clearly visible in Tab.~\ref{tab:metric_DSE}: a complete disregard for \textit{Data-Centric} metrics. These systems are designed to answer, ``Do I trust this user?'' but not the more crucial question for today's information ecosystem: ``Is the content (data) this user is sharing accurate, reliable, or authentic?''. A trustworthy user can still unknowingly share misinformation, a scenario these models are ill-equipped to handle.

Furthermore, as shown in Tab.~\ref{tab:design_DSE}, the reviewed systems do not employ any advanced \textit{Privacy-Preserving Techniques}, a significant concern for a domain built on personal data. For social network TRMSs to evolve into effective tools for trustworthy data sharing and combating misinformation, a paradigm shift is required. Future research can focus on integrating robust \textit{Data-Centric} metrics to directly assess the veracity of the shared content itself, moving beyond just the reputation of the source.

\noindent
\textbf{Data Market:}
Data markets represent the economic culmination of data sharing, creating ecosystems where trust directly underpins financial value. The primary trust challenges in this domain are to accurately reflect the quality of data assets in their valuation and to ensure fair, transparent, and secure transactions between data providers (sellers) and data consumers (buyers). The literature in this area is nascent but highlights a move towards more holistic trust frameworks.

The work by Chowdhury \textit{et al.}~\cite{chowdhury2019trust} exemplifies a purpose-built TRMS for a sensitive data market. As detailed in Tab.~\ref{tab:design_DSE}, it employs a \textit{Centralized} but \textit{Bidirectional} and \textit{Role-Specific} design, which is essential for managing the distinct interactions between health data providers and researchers. Its additional feature, shown in Tab.~\ref{tab:evaluation_DSE}, is its use of a \textit{Combined} evaluation metric with \textit{Hybrid} trust signals. This is a step forward, as it is the only reviewed system in the digital service ecosystems category to formally integrate both \textit{Data-Centric} metrics (\textit{e.g.}, \textit{Authenticity}, \textit{Reliability}) and \textit{Entity-Centric} metrics (\textit{e.g.}, \textit{Compliance}, \textit{Consent}), as detailed in Tab.~\ref{tab:metric_DSE}. In contrast, the study by Hesse \textit{et al.}~\cite{hesse2020takeaway} explores the economic implications of reputation by investigating its transferability between platforms. Its simpler, \textit{Unidirectional} and purely \textit{Entity-Centric} model highlights the challenges in creating portable reputation scores that retain their value across different market contexts.

\textit{Discussion:}
Our analysis shows that TRMSs for dedicated data markets are an emerging but vital research area. The reviewed literature presents a spectrum of maturity, from studies on the economic value of reputation to the design of sophisticated, purpose-built platforms. The framework by Chowdhury \textit{et al.}~\cite{chowdhury2019trust} serves as an important blueprint, as it is the only model identified that begins to bridge the \textit{entity-centric versus data-centric} gap by adopting a \textit{Combined} evaluation that assesses both the data provider and the data.

As shown in Tab.~\ref{tab:security_DSE}, even these market-focused models do not yet comprehensively address key security threats like \textit{Sybil attacks} or \textit{Collusion}. Furthermore, the challenge of creating a standardized, interoperable reputation that is portable across platforms—a problem highlighted by Hesse \textit{et al.}~\cite{hesse2020takeaway}—remains a major barrier to a fluid, global data economy. Future work can focus on maturing these holistic, \textit{Combined} TRMSs, integrating robust ``security by design'' principles, and developing standards for reputation portability to build the trusted and efficient data marketplaces.

\vspace{0.3cm}
\noindent
\textbf{Summary:}
The cross-domain analysis of TRMSs reveals consistent architectural patterns and a persistent research gap. While distributed autonomous systems naturally favor decentralized architectures and digital service ecosystems often rely on centralized or federated control, a unifying theme emerges. The evaluation frameworks are, without exception, overwhelmingly entity-centric, designed to assess the behavior and reliability of the participants. Existing TRMSs consistently overlook two dimensions for trustworthy data sharing: a rigorous, data-centric assessment of the shared asset's quality and the bidirectional evaluation of the data consumer's compliance. This foundational gap motivates the research agenda presented in the following section.

\section{Open issues and future directions}
\label{sec:challenges}
The increasing importance of data sharing across industries underscores the need for reliable, robust, and effective TRMSs; yet, their design is fraught with challenges, as detailed throughout this survey. Although existing surveys~\cite{wang2022heterogeneous,khan2020social,wei2022trust,fotia2023trust} have already highlighted issues like the cold-start problem, heterogeneity, and privacy, this work identifies several open challenges that have been largely overlooked. Building on this analysis, this section outlines these gaps and proposes promising future directions, drawing on insights from emerging technologies to inform the development of next-generation TRMSs.

\noindent
\textbf{Unified and Context-Aware Evaluation Framework.}
A barrier to the advancement and widespread adoption of TRMSs is the lack of a consistent evaluation methodology. The current landscape is fragmented, with systems often relying on domain-specific criteria and custom-built simulations. While effective in isolation, this approach hampers the generalizability, interoperability, and comparative assessment required to mature the field. The development of such a framework can provide immense value to both academia and industry: it enables objective comparison to accelerate innovation, fosters interoperability for portable reputations, and builds practitioner confidence by offering a clear, evidence-based method for selecting the right TRMS.

To overcome these challenges, future research must pursue a dual-pronged research agenda.
The first pillar of this agenda is the development of unified benchmarks. This involves creating a public repository of standardized datasets, both real-world and simulated, that contain labeled examples of trustworthy and malicious behaviors~\cite{bankovic2011detecting} (e.g., bad-mouthing, collusion). Alongside these datasets, a formal set of cross-domain performance metrics must be established~\cite{wu2024privacy}, covering key dimensions such as Accuracy (correctly identifying malicious actors), Robustness (resilience to specific attacks), Scalability (performance under load), and Explainability.
The second pillar must address the fact that no single set of evaluation metrics are universally applicable, thus necessitating context-aware evaluation mechanisms. Current systems often aggregate trust signals without considering their relative significance, leading to scores that lack relevance~\cite{fan2020decentralized}. Future work can address this by employing feature attribution techniques (e.g., SHAP, LIME) to determine the marginal contribution of each signal~\cite{ratul2021evaluating}. This would enable a system to dynamically adjust metric weights based on a ``context vector'' that describes the specific scenario. For instance, the framework would automatically prioritize data privacy compliance in a healthcare setting but shift focus to low latency in a Vehicle-to-Everything (V2X) network, leading to far more precise and meaningful trust evaluations.
The primary research challenges in realizing this vision include establishing and maintaining non-gameable benchmarks and developing computationally efficient methods for real-time contextual adaptation.

\noindent
\textbf{Compliance as a Trust Metric.}
Current TRMSs fail to adequately quantify compliance with complex data usage policies and regulations. While existing semantic tools can identify rules within legal texts, they largely fail to assess the degree of adherence in practice, limiting the accuracy of trust evaluations in regulated environments like those governed by GDPR~\cite{agarwal2018legislative,kirrane2018scalable}. Future research can focus on developing more sophisticated methods to measure and integrate compliance into trust profiles, with LLMs presenting a promising but challenging frontier.

The advanced natural language understanding (NLU) and reasoning capabilities of LLMs make them uniquely suited to bridge this gap~\cite{zhao2023survey}. Future work could involve designing a compliance-aware LLM agent capable of parsing and interpreting the intricate clauses within regulations and DSAs. This agent could then analyze evidence, such as data usage logs or API call records, to automatically detect potential violations. The output would not be a simple binary judgment but a quantifiable compliance score, derived from metrics such as the frequency, severity, and context of violations. For example, a minor data formatting error would carry a lower penalty than the unauthorized sharing of personal data, and the weights of these violations could be adapted to specific domains, such as prioritizing privacy in healthcare or service-level adherence in finance.

However, the integration of LLMs into a trust framework introduces its own risks, most notably model hallucination~\cite{ji2023towards}. A potential avenue for future research is therefore to address the paradox that a technology prone to fabricating information could exacerbate the very trust issues it is meant to solve. An LLM hallucinating a compliance breach could unfairly destroy a provider's reputation, while failing to detect a real violation could compromise the entire ecosystem. Research must therefore focus on developing robust mitigation strategies. These could include implementing retrieval-augmented generation (RAG) frameworks~\cite{lewis2020retrieval} to ensure the LLM's reasoning is strictly grounded in a verified knowledge base of rules and evidence, creating HITL workflows for verifying high-severity violations~\cite{pandey2022modeling}, and advancing explainable AI (XAI) techniques~\cite{dwivedi2023explainable} so the model must cite the specific rule and evidence for its judgment. Ultimately, harnessing LLMs for compliance assessment requires a ``trust but verify'' approach, where research on model safety and reliability is as important as research on its capabilities.

\noindent
\textbf{Data-Centric Trust Evaluation.}
The analysis of existing TRMSs in Sec.~\ref{sec:comparison} makes it clear that they are predominantly entity-centric, designed to answer the question, ``Do I trust the Data Provider?'' While important, this approach overlooks a more fundamental question for data sharing: ``Do I trust the data itself?'' A highly reputable provider, for instance, could inadvertently offer a dataset that is outdated, incomplete, or unfit for a specific purpose. This highlights the need for a necessary shift towards a data-centric trust model, where the characteristics of the data asset (discussed in Sec.~\ref{sec:evaluation metrics}) are treated as first-class metrics in atomic trust evaluation.

This shift towards data-centric evaluation represents another research direction for several key reasons. Firstly, it directly addresses fitness for purpose, since a provider's trustworthiness is irrelevant if the data does not meet the consumer's specific needs. Secondly, it enables contextual trust by recognizing that a dataset's value is not absolute but depends on the application~\cite{mousa2021multi}; for example, data suitable for market analysis may lack the precision required for a medical diagnostic model~\cite{ebrahimi2022quantitative}. Finally, it provides a more robust and objective foundation for trust, as data-centric metrics can be verified through automated analysis, making the system less susceptible to manipulation than those based purely on subjective, entity-level ratings.

To quantify data-centric trust, future research could focus on defining and measuring the data quality metrics discussed in Sec.~\ref{sec:evaluation metrics}. These diverse metrics can be combined into a composite data trust score, a weighted value where the weights are dynamically set by the data consumer's context; for instance, a user needing real-time analytics would prioritize timeliness~\cite{pipino2002data}, while another might prioritize completeness~\cite{pipino2002data,evans2006scaling}. However, implementing this approach poses several challenges. The primary challenge is automating metric measurement at scale, especially for assessing accuracy without a ``golden dataset.'' Furthermore, research is needed to develop standardized formats for provenance metadata and to determine how this new data-centric score can be fused with an entity's reputation for a single, holistic trust assessment.

\noindent
\textbf{Accountability and Incentive Mechanisms.}
One future research direction lies in strengthening the accountability and incentive mechanisms within TRMSs. The effectiveness of any trust model hinges not just on the accuracy of its calculations, but on a foundational framework that enforces responsibility and motivates positive behavior. Many current systems focus heavily on the algorithmic aspects of trust evaluation but neglect the socio-economic dynamics that make a reputation score meaningful. Without clear consequences for untrustworthy actions or tangible benefits for maintaining a high reputation, the system lacks the power to shape participant behavior effectively. Therefore, future work must focus on designing and integrating robust mechanisms that hold entities accountable for their actions while actively incentivizing the trustworthy conduct vital for a healthy data sharing ecosystem. This approach is essential for creating a self-correcting environment that naturally deters malicious actors by raising the cost of attacks, while simultaneously encouraging honest participation by rewarding valuable contributions.

To enhance accountability, research can explore several technical and economic avenues. A key challenge in many systems is the ``Sybil attack'', where malicious actors evade a poor reputation by creating new identities~\cite{zhou2011p2dap}. This can be countered by investigating methods to link trust profiles to verified digital identities, such as Decentralized Identifiers (DIDs)~\cite{fotiou2023self}. Furthermore, direct economic consequences can be implemented through staking and slashing mechanisms, where participants must deposit a financial stake as collateral. If the system verifies a violation, this stake is forfeited, creating a powerful disincentive for malicious behavior. To ensure such penalties are applied fairly, these systems must be underpinned by immutable auditable logs, potentially using blockchain technology to create a tamper-proof record of all transactions and evaluations that can serve as undeniable evidence during dispute resolution.

Complementing these accountability measures, a sophisticated incentive structure is required to motivate proactive, trustworthy conduct. Beyond simple penalties, positive reinforcement can be achieved by treating reputation as a tangible asset itself, where a high score unlocks direct benefits like reduced transaction fees, priority access to premium data, or enhanced voting power in ecosystem governance. This can be further supported by designing tokenomic models that use micropayments to reward participants for providing timely, high-quality feedback that is corroborated by others. Non-financial motivators can also be highly effective; research into gamification can yield systems that use badges, leaderboards, and tiered statuses to recognize and celebrate consistently reliable actors. However, designing these systems presents a delicate balance and several open research questions. Future work must address how to structure incentives to encourage quality over quantity, ensure penalties are applied fairly, create sustainable economic models, and mitigate ``rich-get-richer'' dynamics that could stifle new entrants.

\noindent
\textbf{Explainable Trust Evaluation.}
For a trust evaluation to be effective, its computation must be transparent, its scores clearly justified, and its rationale understandable to all stakeholders. This paradigm, known as explainable trust evaluation, shifts the focus from the trust score itself (the ``what'') to the underlying evidence and logic that produced it (the ``why''). When a system details the metrics it uses—such as interaction history, data quality assessments, or compliance records—and how they contribute to a final score, it provides actionable insights and discourages malicious actors by increasing visibility. Despite this, many current TRMSs function as ``black boxes'', lacking sufficient focus on explainability. This deficiency not only hinders user adoption but also creates vulnerabilities to attacks like bad-mouthing and collusion, where trust scores can be unfairly manipulated without a clear audit trail~\cite{bankovic2011detecting}.

Prioritizing explainability as a core design principle is therefore a key research direction for several reasons. First, it fosters user trust in the TRMS itself, as stakeholders are more likely to rely on scores they can understand. Second, it provides actionable intelligence, allowing entities with declining scores to identify and rectify specific issues, such as poor data timeliness or DSA violations. Third, it enhances robustness against attacks; by requiring evidence-based justifications for ratings, the system can automatically flag anomalous or unsubstantiated feedback, making coordinated collusion much harder to perform. Finally, it provides a transparent foundation for dispute resolution by offering an auditable trail of evidence.

Future research should explore several technical pathways to achieve this. One is developing interpretable-by-design models, such as simple weighted-score systems or rule-based logic, where the link between inputs and outputs is inherently clear. While transparent, these models may lack the accuracy of more complex systems. A second approach involves applying post-hoc explanation techniques (e.g., SHAP, LIME) to complex ``black-box'' models to identify the key features that influenced a score~\cite{ratul2021evaluating}. A third, highly promising direction involves leveraging LLMs~\cite{bilal2025llms}. Research can focus on using LLMs to generate human-readable narratives that synthesize complex data into a clear justification, such as: ``Your score decreased this week primarily due to two `low data quality' ratings and one instance of DSA non-compliance.''

Furthermore, the advanced NLU capabilities of LLMs open another research avenue: unstructured feedback analysis. An LLM could parse text-based user reviews to automatically extract and categorize nuanced reasons for ratings (e.g., distinguishing poor data quality from slow delivery), adding a rich, qualitative layer to an entity's trust profile~\cite{zhang2024logparser}. This same capability can enhance system integrity by automatically auditing reviews to detect fraudulent patterns or coordinated malicious feedback. The primary challenges for future work involve navigating the trade-off between model accuracy and explainability; ensuring explanations faithfully represent the model’s reasoning (i.e., avoiding LLM ``hallucination''); and providing transparency without revealing exploitable vulnerabilities in the trust algorithm itself.

\noindent
\textbf{Security by design in TRMSs.}
Another key research direction is the adoption of a security-by-design philosophy in the development of TRMSs~\cite{casola2020novel}. Many existing models, while sophisticated in their trust calculations, lack robust security architectures, as these considerations are often treated as an afterthought rather than a core requirement. This leaves them vulnerable to the diverse range of attacks outlined in Sec. \ref{sec:threats}. By embedding security principles into every phase of the system lifecycle—from initial architecture to deployment and operation—this proactive approach aims to build TRMSs that are not just accurate, but fundamentally resilient against adversarial attacks.

The TRMS itself must be the most trusted component in a data sharing ecosystem; if its integrity is compromised, the entire foundation of trust collapses. A proactive security posture is essential because TRMSs are high-value targets for attackers seeking to illegitimately boost reputations or slander competitors. Furthermore, for TRMSs to be viable in open and potentially hostile environments like public data marketplaces, they must be designed with the explicit assumption that they will operate among malicious actors. Security by design thus shifts the paradigm from reactively patching vulnerabilities to proactively engineering a system capable of withstanding attacks from the outset.

To implement this methodology, future research can focus on: 
\textit{i) Foundational Threat Modeling:} The first is adapting established methodologies like STRIDE~\cite{khan2017stride} to the unique context of TRMSs. This allows for the identification of threats such as collusion, bad-mouthing, and Sybil attacks before development begins.
\textit{ii) Attack-Resilient Technologies:} The second area is the integration of advanced cryptographic tools and algorithms. This includes leveraging ZKPs for verifiable claims without revealing sensitive data~\cite{aziz2025enhancing}, SMPC for privacy-preserving aggregate calculations~\cite{zhao2019secure}, and blockchain-based immutable ledgers to ensure a tamper-proof, auditable log of all trust-related interactions~\cite{hasan2022privacy}.
\textit{iii) Dynamic and Adaptive Defense:} A third pillar, acknowledging that threats evolve, is the development of adaptive defense mechanisms. This involves using machine learning for real-time anomaly detection to identify suspicious patterns (e.g., a coordinated wave of negative ratings) and trigger automated countermeasures, such as isolating suspicious nodes or requiring a higher burden of proof.
The primary research challenges include managing the significant performance overhead of advanced cryptography, balancing robust security with system usability, and developing methods for the formal verification of a TRMS's resilience against specific classes of attacks.

\section{Conclusion}
\label{sec:conclusion}
In this paper, we present a comprehensive survey of trust and reputation in data sharing. We propose an hierarchical architecture of TRMS in data sharing. Based on the proposed architecture, we introduce novel taxonomies for system design and trust evaluation, highlighting the trustworthiness of data, roles, and processes in data sharing. We expand the evaluation dimensions of data and entities into a unified framework, designed to enhance the generalizability and interoperability of TRMSs. 
Furthermore, we summarize key threats to TRMSs and their implications.
Using the proposed evaluation framework, we conduct a systematic analysis of existing TRMS implementations across various distributed autonomous systems and digital service ecosystems. Our critical review identifies open challenges and suggests leveraging emerging technologies such as RL, LLMs, and blockchain to address these issues. These advancements hold the potential to enhance the explainability, comprehensiveness, and accuracy of TRMSs, fostering secure and efficient data-sharing ecosystems.

\section*{Acknowledgment}
This work was partially funded by the Horizon Europe project \href{https://datapact.eu/}{DATAPACT} (No. 101189771) and the UKRI Horizon Europe guarantee funding scheme for the Horizon Europe projects \href{https://raise-science.eu/}{RAISE} (No. 101058479) and \href{https://www.upcast-project.eu/}{UPCAST} (No. 101093216).

\ifCLASSOPTIONcaptionsoff
  \newpage
\fi

\bibliographystyle{IEEEtran}
\bibliography{reference}

\end{document}